\documentclass[oneside,11pt]{article}

\usepackage{tgpagella}
\usepackage[english]{babel}    
\usepackage{enumerate} 
\usepackage{multirow}
\usepackage{graphics,latexsym,amsfonts}       
\usepackage{amssymb,amsthm,hyperref,mathtools}   
\usepackage[dvipsnames]{xcolor} 
\usepackage{mathrsfs} 
\usepackage{graphicx} 
\usepackage{float} 
\usepackage{ulem}    
\usepackage{picture}
\hypersetup{
    colorlinks, 
    linkcolor={red!50!black},  
    citecolor={blue!50!black},  
    urlcolor={blue!80!black}   
}
\usepackage[longnamesfirst]{natbib}    
\usepackage{tabularx}

\def\reference#1{\href{#1}{Cliquer ici pour voir une r\'ef\'erence.}} 
\usepackage{sectsty} 

\usepackage{pstricks}
\usepackage{pst-grad} % Per sfumature
\usepackage{pst-plot}
\usepackage{pst-node}
\usepackage{pst-text}
\usepackage{setspace}
\usepackage{caption}
\sectionfont{\normalfont\scshape\centering}
\subsectionfont{\centering}
\providecommand{\U}[1]{\protect \rule{.1in}{.1in}}
{\normalfont\itshape}

\usepackage[top=1.25in,bottom=1.25in,left=1.25in]{geometry}
\def\reference#1{\href{#1}{Click to see a reference}} 

\def\X{\mathscr{X}}
\def\es{\varnothing}
\makeatletter

\newtheoremstyle{mytheoremstyle} % name
    {\topsep}                    % Space above
    {\topsep}                    % Space below
    {\itshape}                   % Body font
    {}                           % Indent amount
    {\scshape}                   % Theorem head font
    {.}                          % Punctuation after theorem head
    {.5em}                       % Space after theorem head
    {}  % Theorem head spec (can be left empty, meaning 'normal')

\theoremstyle{mytheoremstyle}
\newtheorem{theorem}{Theorem} % reset theorem numbering for each chapter
\newtheorem*{theorem*}{Theorem}

\newtheorem{lemma}{Lemma}
\newtheorem{corollary}{Corollary} 
\newtheorem*{corollary*}{Corollary} 
\newtheoremstyle{mydefinitionstyle} % name
    {\topsep}                    % Space above
    {\topsep}                    % Space below
    {}                   % Body font
    {}                           % Indent amount
    {\scshape}                   % Theorem head font
    {.}                          % Punctuation after theorem head
    {.5em}                       % Space after theorem head
    {}  % Theorem head spec (can be left empty, meaning "normal"ï)
\theoremstyle{mydefinitionstyle}
\newtheorem{definition}{Definition} % definition numbers are dependent on theorem numbers
\newtheorem{example}{Example} % same for example numbers

\newtheorem*{question*}{Question}

\newtheorem{remark}{Remark}

\makeatletter

\title{\bf Compromise-based Random Utility Models\thanks{The author wishes to thank Gennaro Anastasio, Valentino Dardanoni, Jean-Paul Doignon, Francesco Drago, Paolo Ghirardato, Alfio Giarlotta, M. Ali Khan, Paola Manzini, Marco Mariotti, Daniele Pennesi, Ernesto Savaglio, Lorenzo Stanca, Ester Sudano, Christopher Turansick, and Xinyang Wang for several comments and suggestions.
Special thanks go to Davide Carpentiere, who provided several comments to improve the proofs of the results.
The author received no financial support for the research, authorship, and/or publication of this article.
}}

\author{ 
\textsc{Angelo Enrico Petralia}\thanks{University of Catania, Catania, Italy. angelo.petralia@unict.it}  
}

\usepackage{fancyhdr} 
\date{}

\begin{document}
\sloppy 
\maketitle
\begin{abstract}

{In many choice problems the evaluation of alternatives is determined by a mediation between opposite judgments.
In these situations the decision maker (DM) may not  maximize her true preference, but some compromise on it, in which the first $i$ options are rated according to the adversarial ranking.
Compromise-based Random Utility Models (compromise-based RUMs), which are RUMs whose support is limited to the compromises on some preference, naturally represent the consequences of the trade-off between antithetical criteria on choices.
Compromise-based RUMs are characterized by the existence of a linear order that allows the experimenter to recover choice probabilities from selections over the ground set, or, alternatively, to verify three behavioral axioms.
%An algorithm  detects the DM's mediation, and elicits her unobservable taste that explains the observed choice.  
%The deterministic declination of my pattern has no empirical power, but it allows to define a \textit{degree of self-punishment}, which measures the extent of the denial of pleasure adopted by the DM in her decision.
%We analyze irrational choices that display the lowest degree of self-punishment, and a characterization of them is provided.
%Moreover, I characterize the choice behavior that exhibits the highest degree of self-punishment, and I show that it comprises almost all choices. 
%We also characterize stochastic self-punishment, which collects all the Random Utility Models (RUMs) whose support is restricted to the compromises on some preference.
Necessary and sufficient conditions for a full identification of the DM's preference and her randomization over compromises are singled out.
In all but two cases, there is a unique justification by compromise of data.
Finally, a degree of compromise, which measures the extent of the mediation embraced by the DM in her decision, is characterized.}

\medskip

\medskip

\end{abstract}

\section*{Introduction}

{Choices are often shaped by a conflict between opposing criteria.
Indeed, frequently the decision maker (DM) is caught between her preference and the opposite judgment.
As documented in many studies on self-control and temptation \citep{GulPesendorfer2004,Noor2011}, eating habits are influenced by a compromise between the DM's inclination toward tempting dishes and her desire to pursue a healthy lifestyle.
If temptation is detrimental for the subject, as in the setting of \cite{RavidSteverson2021}, these two rationales are described by inverse orderings of the alternatives.
The same tension occurs between individualism and reciprocity: in experimental settings such as the dictator game, the gift exchange, and public good experiments the maximization of the DM's profit is countered by that of the others.
In response to this constraint, subjects usually reduce their reward to leave some gains for the other participants \citep{FehrCharness2025}.
On such occasions, the DM adopts a utility function that balances personal earnings and social issues, overcoming the contrast between these two antithetical views. 
This principle regulates also guilt aversion \citep{BattigalliDufwenberg2007,EllingsenJohannessonTjottaTorsvik2010,BellemareSebaldSuetens2017} and self-punishment \citep{FrieheHippelSchielke2021}, empirical facts governed by the dissonance between conflicting judgments. 
The trade-off between the DM's preference and its negation offers a natural justification also for the consequences of self-esteem and confidence in choices, discussed by, among others, \cite{ChuangChengChangChiang2013}, and \cite{KoszegiLoewensteinMurooka2022}.
These authors point out that individuals with a low self-confidence might reject risky and desirable alternatives, and favor options bringing less satisfaction, but also demanding lower effort and ability.

{In the patterns mentioned above the DM's choice is usually explained by involved utility functions, depending on several variables framing the observed behavioral bias.
Despite their elegance, these methods may not provide a realistic description of the decision process since, as highlighted by \cite{HarstadSelten2013}, individuals are not efficient maximizers.
Moreover, utilities are often exogenous parameters, and cannot be retrieved from data.}
Thus, I introduce a simple model of choice in which a compromise with the opposite criterion modifies the DM's endogenous preference, by moving the first $i$ alternatives to the bottom of her judgement, in reverse order.
A collection of linear orders generated by this process, and called \textit{compromises on the DM's preference}, describes the different intensities of DM's mediation, and justifies her choice behavior.
Indeed, I define a subclass of \textit{Random Utility Models (RUMs)}, originally proposed by \cite{BlockMarschak1960}, and, starting from \cite{ManskiandMcFadden1981}, widely adopted also in many econometric applications.
RUMs are stochastic choices explained by some randomization over linear orders.
Instead, \textit{compromise-based Random Utility Models (compromise-based RUMs)}, discussed in this note, are RUMs whose support is restricted to the compromises on some preference.
Our method is illustrated in the following examples.}

\begin{example}[\textit{Temptation vs health}]\label{EXMP:temptation}
	Let $X=\{p,f,s\}$ be the set containing pizza ($p$), fettuccine ($f$), and salad ($s$).
	The DM's {unobserved} preference, which enhances tempting food, is described by the linear order $\rhd$ such that $p\,\rhd f\rhd s.$ 
	If she diets, she may disregard the tastiest alternative, and favor { healthier dishes}.
Thus, she  decides according to a compromise $\rhd_1$ on her original preference defined by $f\rhd_1 s \rhd_1 p$, in which the first item, pizza, is moved to the bottom.
If her dietary compliance is even stronger, her judgment could be completely reversed.
In this case, the DM applies in her selection the compromise $\rhd_2\equiv -\rhd$  satisfying $s\rhd_2 f \rhd_2 p,$ which places the first two items, pizza and fettuccine, to the bottom, in reverse order, and salad on top.
{Note that $-\rhd$ ranks the options in $X$ according to their healthiness.} 
Let  $\rho\colon X\times \X\to[0,1]$ be the stochastic choice defined by 
	\begin{center}
		\smallskip
		\begin{tabular}{ccccc}
		\hline
		& $X$ & $pf$ & $ps$ & $fs$\\
		\hline
		$p$ & $0.3$ & $0.3$ & $0.3$ & $0$\\
		\hline
		$f$ & $0.1$ & $0.7$ & $0$& $0.4$\\
		\hline
		$s$ & $0.6$ & $0$ & $0.7$ & $0.6$\\
		\hline
		\end{tabular}
	\smallskip
	\end{center}
{The reader can check that} $\rho$ can be retrieved by the probability distribution $Pr$ over the DM's true preference $\rhd,$ and the compromises $\rhd_1,\rhd_2,$ such that $Pr(\rhd)=0.3,$ $Pr(\rhd_1)=0.1,$ and $Pr(\rhd_2)=0.6.$    
\end{example}

\begin{example}[\textit{Selfishness vs fairness}]\label{EXMP:fairness}
Assume that $X=\{99,75,50\}$ collects the  percentages of an amount of money that a subject in a dictator game can keep for him.
The rest goes to the passive player.	
The dictator would like to obtain as much money as possible, as indicated by her (unobserved) preference $\rhd$ such that $99 \rhd 75 \rhd 50.$
However, if she has some concerns about the equity of the proposal, she may downgrade the possibility of holding almost the whole sum, and prefer the alternatives in which the passive player receives more.
Thus, in these occasions she adopts the compromise $\rhd_1$ such that $ 75 \rhd_1 50 \rhd_1 99,$ in which the most selfish option is the least preferred, but she still gets an advantage from her position.
{If the dictator is extremely sensitive to fairness, she blames any allocation that allows her to gain more than the opponent.
Therefore, her judgment is described by the compromise $\rhd_2\equiv -\rhd$ satisfying $50 \rhd_2 75\rhd_2 99$, and expressing only the equity of the offers.}
 Consider the stochastic choice  $\rho\colon X\times \X\to[0,1]$ defined by  
	\begin{center}
		\smallskip
		\begin{tabular}{ccccc}
		\hline
		& $X$ & $99\;75$ & $99\;50$ & $75\;50$\\
		\hline
		$99$ & $0.5$ & $0.5$ & $0.5$ & $0$\\
		\hline
		$75$ & $0.3$ & $0.5$ & $0$& $0.8$\\
		\hline
		$50$ & $0.2$ & $0$ & $0.5$ & $0.2$\\
		\hline
		
		\end{tabular}
	\smallskip
	\end{center}
 Note that $\rho$ is justified by a probability distribution on all the rankings over $X$ with support containing only the dictator's preference $\rhd$, and its distortions $\rhd_1,$ and $\rhd_2$. 
 Indeed, it is enough to assume that with probabilities $Pr(\rhd)=0.5, Pr(\rhd_1)=0.3, Pr(\rhd_2)=0.2$ the DM's pick in each menu is guided respectively by $\rhd, \rhd_1,$ and $\rhd_2$.    
\end{example}

\begin{example}[\textit{High self-esteem vs low self-esteem}]\label{EXMP:low_self_esteem}
Consider the set $X=\{h,m,l\}$ containing three tasks that respectively offer a high ($h$), medium ($m$), and low ($l$) reward, and proportional levels of individual skills and losses, in case of failure. 
The preference of a confident DM, who aims to obtain the highest prize, is described by the linear order $\rhd$ such that $h\rhd m \rhd l.$
However, a DM who believes she can perform tasks $m$ and $l$, but she cannot successfully finish the task $h$,  may neglect $h$, and base her decision on the compromise $\rhd_1$ satisfying $m\rhd_1 l \rhd_1 h.$
Moreover, if her self-esteem is even lower, she would put on top of her ranking $\rhd_2\equiv -\rhd$ the alternative $l,$ the unique task she can handle, followed by $m$ and $h$, which bring increasing losses, if not accomplished.
The stochastic choice $\rho\colon X\times \X\to[0,1]$  defined by 
	\begin{center}
		\smallskip
		\begin{tabular}{ccccc}
		\hline
		& $X$ & $hm$ & $hl$ & $ml$\\
		\hline
		$h$ & $0.4$ & $0.4$ & $0.4$ & $0$\\
		\hline
		$m$ & $0.2$ & $0.6$ & $0$& $0.6$\\
		\hline
		$l$ & $0.4$ & $0$ & $0.6$ & $0.4$\\
		\hline
		
		\end{tabular}
	\smallskip
	\end{center} 
	is determined by the probability distribution $Pr$ such that $Pr(\rhd)=0.4,$ $Pr(\rhd_1)=0.2,$ and $Pr(\rhd_2)=0.4.$ \footnote{Note that Examples~\ref{EXMP:temptation}, \ref{EXMP:fairness}, and \ref{EXMP:low_self_esteem} can be explained respectively by multi-dimensional preferences, social preferences, and risk preferences.
However, each of these paradigms cannot justify the datasets displayed in the remaining examples.
Thus, my method is richer, and it accounts for a wider variety of phenomena.
Moreover, each example can be easily rephrased using more than three alternatives.}
\end{example}

{Compromise-based RUMs are characterized by the possibility of recovering the dataset from the probabilities of selection from the ground set.
When the DM's preference is exogenous, an alternative characterization relies on three properties of the choice.
A revealed preference procedure allows to easily test this model on data, and infer the DM's taste.    
I determine the necessary and sufficient conditions under which the DM's preference, and the probability weights over the compromises on it are unique.
Finally, I characterize the degree of compromise of a stochastic choice, i.e, a lower bound to the maximal index of the compromises belonging to the support of some randomization that explains data.   

My contribution to the literature is twofold.
First, motivated by the wide range of economics applications, and the relative simplicity of the approach, devoid of any ex-ante behavioral assumption,  I formalize the consequences of compromise between conflicting criteria on choices.
Second, I contribute to the analysis of RUMs, by proposing a specification in which the DM randomizes only among the compromises on her preference.
A detailed comparison between compromise-based RUMs, RUMs, and their subclasses is provided in Section~\ref{SECT:relation_literature}.
    
The paper is organized as follows. Section~\ref{SECT:preliminaries} collects some preliminary notions.
In Section~\ref{SECTION:model} compromise-based RUMs are investigated.
Specifically, in Subsection~\ref{SECTION:model}\ref{SUBSECTION:Characterization} I propose a characterization of this choice behavior.
 Subsection~\ref{SECTION:model}\ref{SUBSECTION:Identification} is devoted to the identification of the DM's preference and randomization over the compromises on it.
In Subsection~\ref{SECTION:model}\ref{SUBSECTION:degree_self_punishment} I define a measure of compromise, and I characterize it.
In Section~\ref{SECT:relation_literature} I compare my approach with the existing subclasses of RUMs. 
Section~\ref{SECT:concluding_remarks} contains some concluding remarks. 
All the proofs have been collected in the Appendix. }

\section{Preliminaries}\label{SECT:preliminaries}
In what follows, $X$ denotes the \textsl{ground set}, a finite nonempty set of \textsl{alternatives}, or \textsl{items}.
A binary relation $\succ$ on $X$ is \textsl{asymmetric} if $x \succ y$ implies $\neg(y \succ x)$, \textsl{transitive} if $x \succ y \succ z$ implies $x \succ z$, and \textsl{complete} if $x \neq y$ implies $x \succ y$ or $y \succ x$ (here $x,y,z$ are arbitrary elements of $X$). 
A \textsl{(strict) linear order} $\rhd$ is an asymmetric, transitive, and complete binary relation.
We denote by $\mathsf{LO}(X)$ the family of all linear orders on $X$.  
{Given $\rhd\in\mathsf{LO}(X)$, denote by $-\rhd\in\mathsf{LO}(X)$ the linear order defined by $x -\rhd\,y$ if $y\rhd x$ for any $x,y\in X$.}
 Any nonempty set $A \subseteq X$ is a \textsl{menu}, and $\X = 2^X \setminus \{\es\}$ denotes the family of all menus.
 Given a linear order $\rhd\in \mathsf{LO}(X)$ and a menu $A \in \X$, the \textsl{maximal alternative of $A$ with respect to $\rhd$}, denoted by  $\max(A,\rhd),$ is the item satisfying $\max(A,\rhd)\in A$, and $	\max(A,\rhd)\rhd y$ for any  $y\in A\setminus\{\max(A,\rhd)\}$.
Instead, the \textsl{minimal  alternative of $A$ with respect to $\rhd$}, denoted by $\min(A,\rhd),$ is the item such that $\min(A,\rhd)\in A$, and $y\rhd \min(A,\rhd)$ for any $y\in A\setminus \{\min(A,\rhd)\}$ hold.

\begin{definition}\label{DEF:stochastic_choice}
A \textsl{stochastic choice function} is a map $\rho:X\times \X \to [0,1]$ such that, for any $A\in\X$, the following conditions hold:
\begin{itemize}
	\item  $\sum_{x\in A}\rho(x,A)=1$, and
	\item $x\not\in A$ implies $\rho(x,A)= 0$. 
\end{itemize}
\end{definition}

The value $\rho(x,A)$ is interpreted as the probability that the item $x$ is selected from the menu $A$.
We refer to a stochastic choice function as a \textsl{stochastic choice}.
Stochastic choices reproduce the outcome of an experimental setting in which the subject performs her selection from each menu multiple times.
Alternatively, they can represent a dataset displaying frequencies of choices implemented by different subjects on the same menus.
We denote by $\Delta(\mathsf{LO}(X))$ the family of all the probability distributions over  $\mathsf{LO}(X)$.
Rationality of stochastic choices is usually defined as follows:
\begin{definition}[\citealt{BlockMarschak1960}]\label{DEF:RUM}
	A stochastic choice $\rho\colon X\times \X\to[0,1]$ is a \textit{Random Utility Model} (for brevity, it is a \textit{RUM}) if there exists a probability distribution $Pr\in \Delta(\mathsf{LO}(X))$ such that for any $A\in\X$ and $x\in A$ 
	$$\rho(x,A)=\sum_{\rhd\in \mathsf{LO}(X)\colon x=\max(A,\rhd)} Pr(\rhd).$$
	We say that $Pr$ \textit{rationalizes} $\rho$. 
\end{definition} 

\section{Compromise-based RUMs}\label{SECTION:model}

We first introduce the notion of \textsl{compromise} on individual preferences.
Before doing so, I need some notation.
Given a set $X$, and some $0\leq i\leq \vert X\vert-1,$ I denote by $X^{\rhd}_i$ the set of the first $i$ items on top of $X$ with respect to $\rhd.$

\begin{definition}\label{DEF:compromise}

Given a set $X$, some $\rhd\in\mathsf{LO}(X)$, and $0\leq i \leq \vert X\vert-1,$ the \textit{i-th compromise on $\rhd$} is the binary relation, denoted by $\rhd_i$, such that
\begin{enumerate}[\rm(i)]
%	\item for any $a,b\in U$, if $a\rhd b$, then $b\rhd_i a$,
	\item for any $a,b\in X\setminus X^{\rhd}_i$, $a\rhd b$ implies $ a\rhd_i b$, and 
	\item for any $a\in X^{\rhd}_i$ and $b\in X$, $a\rhd b$ implies $b\rhd_i a$.
	
\end{enumerate} 
Moreover, a linear order $\rhd^{\prime}\in\mathsf{LO}(X)$ is \textit{a compromise on $\rhd$} if $\rhd^{\prime}\equiv \rhd_i $ for some $i\in~\{0,\cdots,\vert X\vert-1\}.$
 I denote by {$\mathsf{Comp}(\rhd)$} the family $\{\rhd_i\}_{0\leq\, i\,\leq \vert X\vert-1}$ of all the $\vert X\vert$ {compromises on} $\rhd$.
\end{definition}

%\begin{definition}\label{DEF:compromise}
%	
%
%Given a set $X$, and some $\rhd\in\mathsf{LO}(X)$, a binary relation $\rhd_i$ on $X$ is a \textsl{compromise on} $\rhd$ if there is a $\es\subseteq U\subset X$ such that
%\begin{enumerate}[\rm(i)]
%
%	\item for any $a\in U$ and $b\in X$, $b\rhd a$ implies $a \rhd_i b$,
%%	\item for any $a,b\in U$, if $a\rhd b$, then $b\rhd_i a$,
%	\item for any $a\in U$ and $b\in X$, $a\rhd b$ implies $b\rhd_i a$,
%	\item for any $a,b\in X\setminus U$, $a\rhd b$ implies $ a\rhd_i b $, and
%	\item $\vert U\vert=i$.
%
%\end{enumerate} 
%We denote by $\mathsf{Comp}(\rhd)$ the family of all the $\vert X\vert$ compromises on $\rhd$.
%\end{definition}

In the $i$-th compromise $\rhd_i$ of a linear order $\rhd$, the first $0\leq i\leq \vert X \vert-1$ alternatives are shifted, in a reverse order, to the bottom. 
Note that, for any $\rhd\in\mathsf{LO}(X)$ and each $0\leq i\leq \vert X\vert-1$, $\rhd_i$ is a linear order, and it is unique.
Moreover, since $\rhd_{0}\equiv\rhd$, I have that $\rhd\in\mathsf{Comp}(\rhd)$.
Finally, in Definition~\ref{DEF:compromise} I impose that $i<\vert X\vert$,  and I do not include $\rhd_{\vert X\vert},$ since $\rhd_{\vert X\vert}\equiv \rhd_{\vert X\vert-1}\equiv -\rhd.$
{A compromise naturally describes the DM's mediation between her preference $\rhd$ and its negation $-\rhd$.
This mental process ensures that the DM partially respects her preference $\rhd$, as indicated by condition (i) of Definition~\ref{DEF:compromise}, which preserves the relation between items not belonging to $X^{\rhd}_i$.
However, she also adheres to the opposite criterion $-\rhd$, since, due to condition (ii), the items contained in $X^{\rhd}_i$ are downgraded, and occupy in the compromise $\rhd_i$ the same position they hold in $-\rhd.$ 
Thus, the DM reconciles her judgment with the adversarial one, giving priority, to some extent, to the options of $X$ decently placed according to $\rhd,$ but that are not severely penalized by $-\rhd.$
When $i=0,$ the DM follows her true preference.
However, as $i$ approaches $\vert X\vert-1$, the adversarial judgement is progressively favored.
Indeed, condition (ii) implies that the best $i$ alternatives are worse than
any other option, even those that are on top of opposite ranking.
For instance, in Example~\ref{EXMP:temptation}, according to the compromise $\rhd_1$ the option $p$, which is the most tempting food, is deemed worse than $s$, the healthiest one, since it goes way too far in breaking the diet. 
Similarly, if we go back to Example~\ref{EXMP:fairness}, in the compromise $\rhd_2$ the fractions $75$ and $99$, which reveal the DM's selfishness, are inferior to $50$, the only offer guaranteeing a fair allocation.  
The same applies to the risky tasks $h$ and $m$, once, in Example~\ref{EXMP:low_self_esteem}, low self-confidence induces the DM to follow the compromise $\rhd_2$, in which project $l$, bringing no losses, is on top.  
Finally, when $i$ equals $\vert X\vert-1$, the DM experiences the most severe compromise, and embraces the negation of her preference.
Note that Definition~\ref{DEF:compromise} in many cases distinguishes between the DM's true preference and the opposite judgment: the reader can check, for instance, that in Example~\ref{EXMP:temptation} the linear order $\rhd_1$  is a compromise on $\rhd$, but not on $-\rhd$.
Identification is ambiguous only when at most two rankings are considered, as discussed in Subsection~\ref{SUBSECTION:Identification}.  

I now consider a stochastic choice behavior affected by a mediation between the DM's true preference, projected onto a single hedonic dimension, and its negation.
%Indeed, I assume that the DM's true preference is projected onto a single hedonic dimension, with respect to which the observed behavior appears to be self-harming.
Some notation is needed: given a linear order $\rhd\in\mathsf{LO}(X)$, I denote by $\Delta(\mathsf{Comp(\rhd)})$ the family of all probability distributions over the set $\mathsf{Comp}(\rhd)$.

\begin{definition}\label{DEF:Stochastic_lack_of_confidence}
	A stochastic choice $\rho\colon X\times \X\to [0,1]$ is  a  \textit{compromise-based Random Utility Model (compromise-based RUM)} if there is $\rhd\in\mathsf{LO}(X)$ and $Pr\in\Delta(\mathsf{Comp(\rhd)})$ such that
		$$\rho(x,A)=\sum_{\rhd_{i}\in \mathsf{Comp}(\rhd)\colon x=\max(A,\rhd_i)}Pr(\rhd_i)$$
	holds for any $A\in\X$ and $x\in A.$ 
We say that the pair $(\rhd,Pr)$ \textit{justifies by compromise} $\rho$, and it is \textit{a justification by compromise of} $\rho$. 
	Moreover, I denote by $\mathsf{JC}_{\rho}$ the set $\left\{(\rhd,Pr)\in \mathsf{LO}(X)\times \Delta(\mathsf{LO(X)})\,\vert\, (\rhd,Pr)\,\text{justifies by compromise}\,\rho\right\}$.
	\end{definition}

Compromise-based RUMs are RUMs whose support is a subset of the collection $\mathsf{Comp}(\rhd)$ of the compromises on some preference $\rhd\in\mathsf{LO}(X)$, and display the behavior of a DM who applies with some probability a mediation between her taste and the opposite view.
Alternatively, compromise-based RUMs can be interpreted as the outcome of an experiment performed over a population of individuals that share the same preference  over the alternatives, and, when they face a given menu, exhibit different levels of compromise.\footnote{This interpretation is valid if we assume, for instance, that alternatives are monetary payoffs.}
Compromise-based RUMs can be also represented by the DM's randomization over some distortions of her utility.
Before formally presenting this fact, I need some notation.
Given a finite set $\mathcal{T}=\{\alpha\in \mathbb{R}\}$ of real values, I denote by $\Delta(\mathcal{T})$ the family of all the probability distributions on $\mathcal{T}$.
Moreover, I denote by $\mathbf{1}_{\{\,\mathcal{C}\}}$ the indicator function that gives $1$ if condition $\mathcal{C}$ is satisfied, and $0$ otherwise.
The following result holds.

\begin{lemma}\label{LEM:utility_representation_compromise_based_RUMs}
	Given a stochastic choice $\rho\colon X\times \X\to [0,1]$ on $X$, the following are equivalent.
	\begin{enumerate}[\rm(i)]
		\item   the pair
	$(\rhd, Pr)$  is a justification by compromise of $\rho$;
	\item there are an injective function $U\colon X\to \mathbb{R}$ such that 
  $-U(y)<U(z)$ for any $y,z\in X$, a finite set $\mathcal{T}=\{\alpha\in\mathbb{R}\}$, and $P\in \Delta(\mathcal{T})$  such that
		
		$$\rho(x,A)=\sum_{\alpha\in\mathcal{T}\colon x=\max_{y\in A}C_{\alpha}(y)}P(\alpha),$$
		
		 where 
		$$ C_{\alpha}(y)=U(y)\mathbf{1}_{\left\{U(y)< _{\alpha} \right\}}-U(y)\mathbf{1}_{\left\{U(y)\geq _{\alpha}\right\}}$$
		for any $y\in X.$ 
	\end{enumerate}
	  
		Moreover, for any $y,z\in X$ we have that
	\begin{itemize}

		\item[(1)]  $y\rhd z$ if and only $U(y)>U(z)$, and
		\item[(2)] $y\rhd_i z$ for some $\rhd_i\in\mathsf{Comp}(\rhd) $ with $Pr(\rhd_i)>0$ if and only if $C_{\alpha}(y)>C_{\alpha}(z)$ for some $\alpha\in\mathcal{T}$ with $P(\alpha)>0$   
	\end{itemize}
	 Finally, (3) if there are $\rhd_i\in\mathsf{Comp}(\rhd)$ with $Pr(\rhd_i)>0$ and $\mathcal{T}^{\,\prime}\subseteq\mathcal{T}$ such that, for any $\alpha\in T^{\,\prime}$ and for any $y,z\in X$, $P(\alpha)>0$  and we have  $y\rhd_i z$ if and only if  $C_{\alpha}(y)>C_{\alpha}(z)$, then $Pr(\rhd_i)=\sum_{\alpha\in \mathcal{T}^{\prime}} P(\alpha).$
\end{lemma}

The lemma above shows that a compromise-based RUM is determined by a probability distribution over a collection $\mathcal{T}$ of parameters, each of them describing a utility threshold that cannot be reached without unduly overlooking the opposite assessment.
Each $\alpha\in\mathcal{T}$ generates a compromise $C_{\alpha}$ between the DM's utility function $U$ and its negation $-U$, in which all the options whose value exceeds $\alpha$ are assessed according to $-U$.    
Note that, since $-U(y)<U(z)$ holds for any $y,z\in X$, any alternative bringing a utility higher or equal than $\alpha$ follows, according to $C_{\alpha}$, all the options with a valuation below $\alpha$.
Moreover, $U$ represents the DM's preference, each $C_{\alpha}$ represents some compromise on it adopted with positive probability, and $P$ reproduces the DM's randomization.

\begin{remark}
In the proposed framework options are elements of a generic finite set $X$.
However, one can modify and/or add more structure to the domain of alternatives to deal with applications.
For instance, Example~\ref{EXMP:fairness} can be rephrased by assuming that the  set of alternatives is $X=\left\{(x,M-x)\in\mathbb{R}_{+}^{2}\,\vert\, x,M>0 \wedge M-x\leq x \leq M \right\},$ where $M$ is the sum that can be split between the two players, $x$ describes the amount that the DM aims to retain, and $M-x$ is left to the opponent once her offer is accepted.
Any proposal in which the DM receives less than the other subject is excluded from $X$ since it is not regarded as fair by her.
I  also require that the subjects share the same utility $u(x)=x$ over the money they obtain after the game is concluded.
Consider the DM's evaluation $C_{\alpha}$ of the alternatives defined by

$$C_{\alpha}(x,M-x)=u(x) {\mathbf{1}}_{\left\{u(x)<\alpha \right\}}+u(M-x)  {\mathbf{1}}_{\left\{u(x)\geq \alpha \right\}}=x\,{\mathbf{1}}_{\left\{x<\alpha \right\}}+ (M-x)\,{\mathbf{1}}_{\left\{x\geq \alpha \right\}},$$

with $\alpha\in\mathbb{R}$.
If the money obtained by the DM is lower than $\alpha$, then the DM values the proposals looking only at her profit.
However, any allocation that provides the DM with a wealth greater or equal than $\alpha$ is judged according to a fairness criterion, i.e., the sum received by the opponent.   
When $\alpha > M$, alternatives in $X$ are ranked following the linear order $\rhd\equiv\,>$ defined by $$(x,M-x)\rhd (y,M-y)\Longleftrightarrow C_{\alpha}(x,M-x)=u(x)=x>y=u(y)=C_{\alpha}(y,M-y),$$
reflecting her selfish assessment.
When $\alpha\leq M,$ options are ordered by $\rhd_{\alpha}$ defined by $$(x,M-x)\rhd_{\alpha} (y,M-y)\Longleftrightarrow x\,{\mathbf{1}}_{\left\{x<\alpha \right\}}+ (M-x)\,{\mathbf{1}}_{\left\{x\geq \alpha \right\}}> y\,{\mathbf{1}}_{\left\{y<\alpha \right\}}+ (M-y)\,{\mathbf{1}}_{\left\{y\geq \alpha \right\}},$$ i.e.,  either $x,y<\alpha$ and $x>y$ or $y\geq \alpha$ and $y>x$.
Thus, $\rhd_{\alpha}$ is, in the same fashion of Definition~\ref{DEF:compromise}, the \textit{$\alpha$-th compromise on} $>$, in which all the offers bringing to the DM an amount greater or equal than $\alpha$ are moved to the bottom of her preference, in reverse order.
Choice probabilities on each menu of finite cardinality are recovered by a discrete probability distribution with finite support over such compromises.
For instance, if $M=100$, and we consider all the menus which are subsets of $X^{\prime}=\{(99,1),(75,25),(50,50)\}\subseteq X$, it is easy to verify that the dataset $\rho$ presented in Example~\ref{EXMP:fairness} is justified by the family $\{>,\rhd_{99},\rhd_{75}\}$, and the probability distribution $Pr$ such that $Pr(>)=0.5$, $Pr(\rhd_{75})=0.3$, and $Pr(\rhd_{50})=0.2.$\footnote{If menus are intervals, then choice frequencies could be retrieved by a cumulative distribution determined by some density function over compromises.}   
To fit the representation of the model proposed in condition (ii) of Lemma~\ref{LEM:utility_representation_compromise_based_RUMs}, it is enough to suitably modify $C_{\alpha}$ by adding a further cost of exceeding $\alpha$\,:

$$C^{\,\prime}_{\alpha}(x,M-x)=u(x) {\mathbf{1}}_{\left\{u(x)<\alpha \right\}}+[u(M-x)-u(M)]  {\mathbf{1}}_{\left\{u(x)\geq \alpha \right\}}=x\,{\mathbf{1}}_{\left\{x<\alpha \right\}}-x\,{\mathbf{1}}_{\left\{x\geq \alpha \right\}}.$$
\end{remark}

Since compromise-based RUMs are a subclass of RUMs, the model presented in Definition~\ref{DEF:Stochastic_lack_of_confidence} and in Lemma~\ref{LEM:utility_representation_compromise_based_RUMs} is testable, and it can be characterized, as showed in the next subsection.
}

\subsection{Characterization}\label{SUBSECTION:Characterization}

Before providing a characterization of compromise-based RUMs, I discuss some necessary conditions of them, which allow to detect the DM's {mediation} from data. 
First, I need some preliminary notation, partially presented in the proof of Lemma~\ref{LEM:utility_representation_compromise_based_RUMs}, and a key result.   
Order the ground set $X$ as $\left\{x^{\rhd}_1,\cdots,x^{\rhd}_{\vert X\vert}\right\},$ where $x^{\rhd}_i\rhd x^{\rhd}_j$ if and only if $i<j.$
Thus, given some $1\leq j\leq \vert X\vert$, $x_j^{\rhd}$ denotes the $j$-th item of $X$ with respect to $\rhd$.
Moreover, denote by $x_j^{\uparrow\rhd}$ the set $\{y\in X \colon y\,\rhd x^{\rhd}_j \}=\{x^{\rhd}_h\in X\colon h< j\}$, by $x_j^{\downarrow\rhd}$ the set $\{y\in X \colon x^{\rhd}_j\rhd y \}=\{x^{\rhd}_k\in X\colon k> j\},$ by $A_{x_j^{\uparrow\rhd}}$ the set $\left(x_j^{\uparrow\rhd}\cap \,A\right)$, and by $A_{x_j^{\downarrow\rhd}}$ the set $\left(x_j^{\downarrow\rhd}\cap \,A\right)$. 
We have:

\begin{lemma}\label{LEM:equality_definitions}
	For any $\rhd\in\mathsf{LO}(X)$, any $Pr\in\Delta(\mathsf{Comp(\rhd)})$, any $A\in\X$, and any $x\in A$ such that $x=x_j^{\rhd}$ for some $1\leq j\leq \vert X\vert $, I have that 
	\begin{align*}
	\sum_{\rhd_{i}\in \mathsf{Comp}(\rhd)\colon x=\max(A,\rhd_i)}Pr(\rhd_i)=&\sum_{k\leq j-1} Pr(\rhd_{k})-\mathbf{1}_{\left\{A_{x_j^{\uparrow\rhd}}\neq \es\right\}}\sum_{k< g\colon x^{\rhd}_g=\min\left(A_{x_j^{\uparrow\rhd}},\,\rhd\right)}Pr(\rhd_k)\\
	&+\mathbf{1}_{\left\{A_{x_j^{\downarrow\rhd}}=\es\right\}}\sum_{k\geq j} Pr(\rhd_{k}).
	\end{align*}
\end{lemma}

Lemma~\ref{LEM:equality_definitions} is a computational tool that allows to equivalently define compromised-based RUMs by using indices of the compromises on the DM's true preference. 

 \begin{corollary}\label{COR:nec_cond_compromise-based_choices_general}
A stochastic choice $\rho\colon X\times \X\to [0,1]$ is justified by compromise  by some pair $(\rhd,Pr)$ if and only if

	$$\rho(x^{\rhd}_j,A)=\sum_{k\leq j-1} Pr(\rhd_{k})-\mathbf{1}_{\left\{A_{x_j^{\uparrow\rhd}}\neq \es\right\}}\sum_{k< g\colon x^{\rhd}_g=\min\left(A_{x_j^{\uparrow\rhd}},\,\rhd\right)}Pr(\rhd_k)
	+\mathbf{1}_{\left\{A_{x_j^{\downarrow\rhd}}=\es\right\}}\sum_{k\geq j} Pr(\rhd_{k})$$
	
	for any  $A\in\X$, and any $1\leq j\leq \vert X\vert$ such that $x^\rhd_j\in A$.
\end{corollary}

Corollary~\ref{COR:nec_cond_compromise-based_choices_general} shows that, if a choice is rationalized by compromise by some pair $(\rhd,Pr)$, then the probability of selecting a given item $x^{\rhd}_{j}$, which holds the $j$-th position in her true preference, from a menu $A$, is the sum of two components.
The first is the sum of the probabilities, according to $Pr$, of each compromise $\rhd_k$, with $k\leq j-1$, for which there is no $x^{\rhd}_{h}$, preferred to $x^{\rhd}_{j}$ according to $\rhd$, and contained in $A$, that it is still ranked over $x^{\rhd}_j$ according to $\rhd_k$.
The second component is the sum of the probabilities of any compromise $\rhd_k$, with $k\geq j$, conditioned to absence in the menu of some $x^{\rhd}_l$ in $A$ worse than $x^{\rhd}_j$ according to $\rhd$.

Corollary~\ref{COR:nec_cond_compromise-based_choices_general} implies that if a stochastic choice is a compromise-based RUM, then the probability that the DM has been used in her decision a given compromise on her preference can be easily detected from the dataset.

\begin{corollary}\label{COR:identification_distortions_probability}
	If $\rho\colon X\times \X\to [0,1]$ is justified by compromise by some pair $(\rhd,Pr)$,
	then $Pr(\rhd_{i})=\rho\left(x_{i+1}^{\rhd},X\right)$ for any $0\leq i \leq \vert X \vert -1$.
\end{corollary}

Corollary~\ref{COR:identification_distortions_probability} states the probability that the DM adopted the compromise $\rhd_{i}$ in each selection equals the probability of choosing the item $x^{\rhd}_{i+1}$ from $X$.
We now introduce a property that reveals the inner structure of compromise-based RUMs.

\begin{definition}\label{DEF:ordered_composition}
	A stochastic choice $\rho\colon X\times \X\to[0,1] $ \textit{has an ordered composition} if there is a linear order $\rhd$ on $X$ such that 
	$$\rho(x^{\rhd}_j,A)= \sum_{k\leq j} \rho(x^{\rhd}_{k},X)-\mathbf{1}_{\left\{A_{x_j^{\uparrow\rhd}}\neq \es\right\}}\sum_{k\leq g\colon x^{\rhd}_g=\min\left(A_{x_j^{\uparrow\rhd}},\,\rhd\right)}\rho(x^{\rhd}_k,X)+ 
	\mathbf{1}_{\left\{A_{x_j^{\downarrow\rhd}}=\es\right\}}\sum_{k> j} \rho(x^{\rhd}_{k},X)$$
	
	for any  $A\in\X$, and any $1\leq j\leq \vert X\vert$ such that $x^{\rhd}_j\in A$.
	We say that $\rhd$ \textit{composes} $\rho$.
	
\end{definition}

Definition~\ref{DEF:ordered_composition} requires the existence of a ranking over the alternatives that allows the experimenter to recover choice probabilities from the DM's selection on ground set.
Indeed, the probability of selecting from a menu $A$ an item $x^{\rhd}_j$ holding the $j$-th position in $X$ with respect to $\rhd$ is the sum of two components.
The first member is the sum of the probabilities of picking from $X$ any $x^{\rhd}_{h}$, which comes before $x^{\rhd}_j$ according to $\rhd$, but it is not contained in $A$, and it is preceded, according to $\rhd$, by the minimal item among those that precede $x^{\rhd}_j$ and are contained in $A$.  
The second component is the sum of the probabilities of selecting from $X$ each item $x^{\rhd}_{l}$ that comes after $x^{\rhd}_j$ according to $\rhd$, conditioned to the absence in $A$ of any item that follows $x^{\rhd}_j$.
Compromise-based RUMs are characterized by ordered compositions.
      
\begin{theorem}\label{THM:compromise-based_stochastic_choices_characterization} 
A stochastic choice is a compromise-based RUM 
 if and only if it has an ordered composition.
		\end{theorem}

{Theorem~\ref{THM:compromise-based_stochastic_choices_characterization} shows that the experimenter can check whether a stochastic choice $\rho$ is a compromise-based RUM by verifying that the dataset has an ordered composition.
As for \cite{ApesteguiaBallesterLu2017}, the axiomatization of this model relies on the existence a linear order over the alternatives that determines some regularities in the dataset.\,\footnote{A (negative) existential condition characterizes also uniquely identified RUMs, studied by \cite{Turansick2022}.}
Indeed, in Section~\ref{SECT:relation_literature} I will show that compromise-based RUMs are a subclass of the patterns described by the authors.
Unlike my approach however, in their setting such linear order is fixed \textit{a priori}, allowing for a characterization that offers some behavioral insights about the phenomenon studied by them, and it is not merely combinatorial.
Thus, it seems natural to propose a more interpretable axiomatization of compromise-based RUMs anchored to the DM's preference, which is now considered an exogenous parameter.
To that end, I present the following properties.\footnote{I thank an anonymous referee for such suggesting them.}

\begin{definition}\label{DEF:characterizing_behavioral_properties}
	Let $\rho\colon X\times \X\to [0,1] $ and $\rhd\in\mathsf{LO}(X)$ be respectively a stochastic choice and a linear order on $X$.
	We say that $\rho$ satisfies
	\begin{enumerate}[-]
	\item \textit{persistent compromise} if $\rho(y,A\setminus x)=\rho(y,A)+\rho(x,A)$ for any $A\in \X$ and $x,y\in A$ such that $x\rhd y$ and there is no $z\in A$ satisfying $x\rhd z\rhd y$,
	\item \textit{minimal rewards} if $\rho(y,A\setminus x)=\rho(y,A)+\rho(x,A)$ for any  $A\in \X$ and $x,y\in A$ such that $x=\min(A,\rhd)$ and there is no $z\in A$ satisfying $y\rhd z\rhd x$, and
	\item \textit{compromise-based centrality} if $\rho(y,A\setminus x)=\rho(y,A)$ for any $A\in\X $ and $x,y\in A$ satisfying at least one of the following conditions: 
	\begin{enumerate}
		\item[(1)] there is $z\in A$ such that $y\rhd x\rhd z$, or 
\item[(2)] there is $z\in A$ such that $y\rhd z\rhd x$, or
\item[(3)] there is $z\in A$ such that $x\rhd z\rhd y$. 
\end{enumerate}
 %	\begin{itemize}
%		\item[(a)] either $y\rhd x$ or there is $z\in A$ satisfying $x\rhd z\rhd y$, and
%		\item[(b)] $x\rhd w$ for some $w\in A$ or there is $z\in A$ satisfying $y\rhd z\rhd x$
% 	\end{itemize}  
	\end{enumerate} 
\end{definition}

Persistent compromise requires that if the DM partially favors the opposite judgement $-\rhd$ and chooses $y$, which is dominated (according to $\rhd$) by $x$ without any intermediate option, then, once $x$ is removed from the menu, she will continue to prioritize $-\rhd$, increasing the probability that $y$ is chosen.
Minimal rewards ensures that if $x$, the worst item in the menu, is selected because the DM gives priority to $-\rhd$, then when $x$ is taken off, she is more likely to select $y$, the available alternative providing the lowest utility.
Finally, compromise-based centrality imposes that if there is an available alternative, distinct from $y$, assuming an intermediate value, then, once an item $x$ is removed from the menu, the choice frequency of $y$ does not change.
These three mutually independent properties characterize, if the DM's preference is known, compromise-based RUMs.

\begin{theorem}\label{THM:characterization_compromise_based_RUMs_exogenous_preference}
	Given a stochastic choice $\rho\colon X\times \X\to [0,1]$ on $X$, and $\rhd\in\mathsf{LO}(X)$, the following are equivalent:
\begin{enumerate}[\rm(i)]
	\item there is $Pr\in\Delta(\mathsf{Comp}(\rhd))$ such that $(\rhd,Pr)$ is a justification by compromise of $\rho$;
	\item $\rhd$ composes $\rho$;
	\item $\rho$ satisfies persistent compromise, minimal rewards, and compromise-based centrality.
\end{enumerate}
\end{theorem}}

{
Theorem~\ref{THM:characterization_compromise_based_RUMs_exogenous_preference} states that, once the DM's taste is disclosed, the experimenter can determine whether her choice is driven by compromise by checking the axioms listed in Definition~\ref{DEF:characterizing_behavioral_properties}.
This result relies on the elicitation of the DM's preference.
Thus, in the following subsection I elaborate on the identification strategies.} 

\subsection{Identification}\label{SUBSECTION:Identification}

The proof of Theorem~\ref{THM:compromise-based_stochastic_choices_characterization} reveals that the linear order that composes the dataset is also the DM's preference, and it allows to retrieve the probability distribution over its compromises.
Moreover, Corollary~\ref{COR:identification_distortions_probability} implies that if a pair $(\rhd,Pr)$ justifies by compromise choice data, then $Pr$ is uniquely determined.  
We formalize these insights in the next result. Some preliminary notation: given a stochastic choice $\rho\colon X\times \X\to [0,1]$ on $X$ and a linear order $\rhd\in\mathsf{LO}(X)$, let $Pr_{\rho,\rhd}\in\Delta(\mathsf{Comp}(\rhd))$ be the probability distribution defined by $Pr_{\rho,\rhd}(\rhd_i)=\rho\left(x^{\rhd}_{i+1},X\right)$ for any $0\leq i\leq \vert X\vert-1.$ 
We have:

\begin{corollary}\label{COR:derivation_probability_from_dataset}
If $(\rhd,Pr)$ is a justification by compromise of $\rho$, then $Pr\equiv Pr_{\rho,\rhd}$, and $\rhd$ composes $\rho$.
	If $\rhd$ composes $\rho\colon X\times\X\to[0,1]$, then $(\rhd,Pr_{\rho,\rhd})$ is a justification by compromise of $\rho$.
\end{corollary}

Corollary~\ref{COR:derivation_probability_from_dataset} states that the probability distribution $Pr_{\rho,\rhd}$ is the unique one that, paired with $\rhd$, justifies by compromise the dataset. 
Conversely, once the experimenter finds a linear order $\rhd$ satisfying  the condition of Definition~\ref{DEF:ordered_composition}, he can deduce that the pair $(\rhd,Pr_{\rho,\rhd})$ justifies by compromise $\rho$.
The search of a suitable linear order is not involved for a relatively small number of alternatives, but it may become computationally heavy when the size of the ground set increases.
Indeed, when $\vert X\vert=n$, there are $n!$ linear orders on $X$ that should be examined to verify that the choice has a linear composition.
However, there is a simple technique to check that the dataset satisfies ordered composition, and retrieve some DM's preference that fits data.
\begin{definition}\label{DEF:revealed_preference_algorithm_self_punishment}
	Let  $\rho\colon X\times \X\to[0,1]$ be a stochastic choice $\rho\colon X\times \X\to[0,1]$ on a set of cardinality $\vert X\vert\geq 3.$
	We call the \textit{{compromise-based} revealed preference algorithm} the following procedure:
\begin{itemize}
		\item[$1.$] find $w\in X$ such that $Pr(w,A)=Pr(w,X)$ for any $A\in \X$ such that $\vert A\vert \geq 2$, and $w\in A,$ and set $w=x^{\rhd}_{1};$
		\item[$2.$]  find $x\in X\setminus x^{\rhd}_{1}$ such that
		\begin{itemize}
		\item $Pr(x,A)=Pr(x^{\rhd}_{1},X)+Pr(x,X)$ for any $A\in\X$ such that $\vert A\vert\geq 2,$ $x\in A$, and  $x^{\rhd}_{1}\not\in A,$ 
		\item $Pr(x,A)=Pr(x,X)$ for any $A\in\X$ such that $\vert A\vert\geq 3,$ $x\in A$ and  $x^{\rhd}_{1}\in A,$
		\item  $Pr(x,xx^{\rhd}_{1})=1-Pr(x^{\rhd}_{1},X),$
		\end{itemize}

and set $x=x^{\rhd}_{2};$
\item[$\vdots$]
\item[$j.$] find $y\in X\setminus x^{\rhd}_{1}x^{\rhd}_2\cdots x^{\rhd}_{j-1}$ such that
\begin{itemize}
\item $Pr(y,A)=\sum^{j-1}_{k=1} Pr(x^{\rhd}_{k},X)+Pr(y,X)$ for any $A\in\X$ such that $\vert A\vert \geq 2,$ $y\in A$, and $x^{\rhd}_{1}x^{\rhd}_2\cdots x^{\rhd}_{j-1}\cap A=\es,$ 
\item $Pr(y,A)=\sum^{k=j-1}_{k=g+1}Pr(x^{\rhd}_{k},X)+Pr(y,X),$ for any $A\in\X$ such that $\vert A\vert\geq 3,$ $y\in A,$ $x^{\rhd}_{1}x^{\rhd}_2\cdots x^{\rhd}_{j-1}\cap A\neq \es,$ $x_g=\min (x^{\rhd}_{1}x^{\rhd}_2\cdots x^{\rhd}_{j-1}\cap A,\rhd),$ and $z\in A$ for some $z\in X\setminus  x^{\rhd}_{1}x^{\rhd}_2\cdots x^{\rhd}_{j-1}y,$
\item {$Pr(y,A)=1-\sum^{k=g}_{k=1}Pr(x^{\rhd}_{k},X)$ }for any     
 $A\in\X$ such that $\vert A\vert\geq 2,$ $y\in A,$ $x^{\rhd}_{1}x^{\rhd}_2\cdots x^{\rhd}_{j-1}\cap A\neq \es,$ $x_g=\min (x^{\rhd}_{1}x^{\rhd}_2\cdots x^{\rhd}_{j-1}\cap A,\rhd),$ and $z\not\in A,$ for any $z\in X\setminus x^{\rhd}_{1}x^{\rhd}_2\cdots x^{\rhd}_{j-1}y,$
 \end{itemize}
 and set $y=x^{\rhd}_j;$
\item[$\vdots$]
\item[$\vert X\vert.$]
 Verify that, for $z\in X\setminus x^{\rhd}_{1}x^{\rhd}_2\cdots x^{\rhd}_{\vert X\vert-1}$, {$Pr(z,A)=1-\sum^{k=g}_{k=1}Pr(x^{\rhd}_{k},X)$} for any     
 $A\in\X$ such that $\vert A\vert\geq 2,$ $z\in A,$ and $x_g=\min (x^{\rhd}_{1}x^{\rhd}_2\cdots x^{\rhd}_{\vert X\vert-1}\cap A,\rhd).$ If so, set $z=x^{\rhd}_{\vert X\vert}.$
	\end{itemize}
	
\end{definition}

The following result holds. 

\begin{corollary}\label{COR:algorithm_ordered_composition}
	A stochastic choice $\rho\colon X\times \X\to[0,1]$ on a set of cardinality $\vert X\vert\geq 3$ has an ordered composition if and only if the {compromise-based} revealed preference algorithm can be completed.
	Moreover, a linear order $\rhd\in\mathsf{LO}(X)$ obtained from the {compromise-based} revealed preference algorithm composes $\rho.$
{Finally, given some $\rhd\in\mathsf{LO}(X)$ obtained from the {compromise-based} revealed preference algorithm, $\rho$ satisfies persistent compromise, minimal rewards, and compromise-based centrality.}	
\end{corollary}

Corollary~\ref{COR:algorithm_ordered_composition} states that, using the {compromise-based} revealed preference algorithm, the experimenter  can test whether a stochastic choice has an ordered composition, and, by Theorem~\ref{THM:compromise-based_stochastic_choices_characterization}, {it} is a compromise-based RUM.
{Moreover, due to Theorem~\ref{THM:characterization_compromise_based_RUMs_exogenous_preference}, it satisfies persistent compromise, minimal rewards, and compromise-based centrality.}
Finally, Corollary~\ref{COR:derivation_probability_from_dataset} ensures that, once completed, the procedure described in Definition~\ref{DEF:revealed_preference_algorithm_self_punishment} provides a DM's preference and the randomization among its compromises  {justifying} the observed choice.
An application of {such routine} is provided in the following example.

\begin{example}
Consider the stochastic choice $\rho$ defined in Example~\ref{EXMP:temptation}.
We perform the steps described in Definition~\ref{DEF:revealed_preference_algorithm_self_punishment}.
\begin{itemize}
	\item[$1.$] $Pr(p,X)=Pr(p,pf)=Pr(p,ps)=0.3.$
	Thus, I set $p=x^{\rhd}_1.$
	\item[$2.$] 
	\begin{itemize}
\item $Pr(f,fs)=0.4=0.3+0.1=Pr(p,X)+Pr(f,X),$ and 
\item $Pr(f,pf)=0.7=1-0.3=1-Pr(p,X).$

\end{itemize}
We set $f=x^{\rhd}_2.$
\item[$3.$] $Pr(s,X)=0.6=1-0.3-0.1=1-Pr(p,X)-Pr(f,X),$ $Pr(s,ps)=0.7=1-0.3=1-Pr(p,X),$ and $Pr(s,fs)=0.6=1-0.3-0.1=1-Pr(p,X)-Pr(f,X).$
We set $s=x^{\rhd}_3.$
\end{itemize}   
The revealed preference algorithm can be completed.
By Corollary~\ref{COR:algorithm_ordered_composition} $\rho$ has a ordered composition, the linear order $p\rhd f\rhd s$ composes $\rho$, and {$\rho$ satisfies persistent compromise, minimal rewards, and compromise-based centrality}.
Corollary~\ref{COR:derivation_probability_from_dataset} implies that $(\rhd,Pr_{\rho,\rhd}),$ with $Pr(\rhd)=\rho(x^{\rhd}_1,X)=0.3, Pr(\rhd_1)=\rho(x^{\rhd}_2,X)=0.1,$ and $Pr(\rhd_2)=\rho(x^{\rhd}_3,X)=0.6,$ is a justification by compromise of $\rho$.

\end{example}

The technique exhibited above is easy to implement, and it can be adopted to test {compromise} on stochastic choices defined on ground sets of larger size.
{This procedure can be extended to deal also with noisy data.
Indeed, if there are no linear orders satisfying the algorithm, one can take those minimizing some function of the discards between the frequencies involved in the equalities listed in Definition~\ref{DEF:revealed_preference_algorithm_self_punishment}.
Alternatively, a statistical test for compromise-based RUMs can be obtained by restricting the domain of the RUMs estimators proposed by \cite{AndrewsSoares2010}, \cite{KitamuraStoye2018}, and \cite{ForcinaDardanoni2024} only to the compromises on some preference.}

One may ask whether the revealed preference algorithm  generates a unique ranking among the alternatives, and there is only one justification by {compromise} of {the dataset}.
 Before addressing this issue, it is worth noting that compromise-based RUMs are uniquely identified RUMs.  

\begin{lemma}\label{LEM:uniqueness_up_to_probability_distribution_preliminary}
	If $\rho\colon X\times \X\to [0,1]$ is a compromise-based RUM, then there is a unique $Pr\in\mathsf{LO}(X)$ that rationalizes $\rho.$
\end{lemma}

A consequence of Lemma \ref{LEM:uniqueness_up_to_probability_distribution_preliminary}  is the following.

\begin{corollary}\label{LEM:uniqueness_up_to_probability_distribution}	Assume that there is $\rhd\in{\mathsf{LO}}(X)$ such that $(\rhd,Pr_{\rho,\rhd})$ justifies by compromise $\rho$.
	The following are equivalent for any $\rhd^{\prime}\in\mathsf{LO}(X)$:
	\begin{enumerate}[\rm(i)]
		\item $(\rhd^{\prime},Pr_{\rho,\rhd^{\prime}})$ justifies by self punishment $\rho$ ;
		\item $\{\rhd_i\in\mathsf{Comp}(\rhd)\,\vert\, Pr_{\rho,\rhd}(\rhd_i)>0\}\subseteq \mathsf{Comp}(\rhd^{\prime}).$
	\end{enumerate}
\end{corollary}

Corollary~\ref{LEM:uniqueness_up_to_probability_distribution} suggests that the elicitation of a unique DM's preference, and the associated compromises involved in her randomization may not always be allowed.
Indeed, multiple justifications by compromise exist if the linear orders belonging to the support of the probability distribution that rationalizes a compromise-based RUM belong to the collections of compromises on different preferences.  
To see this, I exhibit in the following example a choice dataset that admits two distinct justifications by compromise.

\begin{example}\label{EXMP:Stochastic_choice_multiple_explanations}
	Let $X=\{w,x,y,z\}$ and $\rho\colon X\times\X\to[0,1]$ be the stochastic choice defined as follows:
	\medskip
	\begin{center}
\begin{tabular}{ ccccccccccccc} 
\hline
   & $X$  & $wxy$ & $wyz$ & $wxz$ & $xyz$ & $wx$ & $wy$ & $wz$ & $xy$ & $xz$ & $yz$ \\ 
\hline
$w$ & $0$& $0$& $0$& $0$& $0$& $0$& $0$& $0$& $0$& $0$& $0$  \\ 
\hline
$x$ & $0.5$ & $0.5$ &$0$ & $0.5$ &  $0.5$ & $1$ & $0$ & $0$ & $0.5$ &$0.5$ & $0$\\ 
\hline
$y$ & $0$ & $0.5$ & $0.5$ & $0$ & $0$ & $0$ & $1$ &$0$ & $0.5$ & $0$ & $0.5$\\
\hline
$z$ &$0.5$ & $0$ & $0.5$ & $0.5$ & $0.5$ & $0$ & $0$ & $1$ & $0$ & $0.5$ & $0.5$ \\
\hline
\end{tabular}
\end{center}
\medskip	
The revealed preference algorithm indicates that the linear orders $\rhd,\rhd^{\prime}\in\mathsf{LO}(X)$ such that {$w\rhd x\rhd y\rhd z$}, and {$w\rhd^{\prime}z \rhd^{\prime} y \rhd^{\prime} x$} compose $\rho.$
By Corollary~\ref{COR:derivation_probability_from_dataset} the pairs $(\rhd,Pr_{\rho,\rhd})$ and $(\rhd^{\prime},Pr_{\rho,\rhd^{\prime}})$ justify by compromise $\rho$.
Note that $Pr_{\rho,\rhd}(\rhd_1)=Pr_{\rho,\rhd}(\rhd_3)=Pr_{\rho,\rhd^{\prime}}(\rhd^{\prime}_1)=Pr_{\rho,\rhd^{\prime}}(\rhd^{\prime}_3)=0.5$.
Moreover, $\rho$ is a uniquely identified RUM, rationalized only by the distribution $Pr\in\Delta(\mathsf{LO}(X))$ such that $Pr(\rhd_1\equiv\rhd^{\prime}_3)=0.5,$ and $Pr(\rhd_3\equiv\rhd^{\prime}_1)=0.5.$ 
 \end{example}

However, a unique justification by compromise is guaranteed by some properties of the dataset, which are displayed in the next result.
We need some preliminary notation.
Given a stochastic choice $\rho\colon X\times\X\to[0,1]$, let $ X^*$ be the set $\{x\in X\vert\,\rho(x,X)> 0\}$.
\begin{theorem}\label{THM:stochastic_self_punishment_identification}
	Let $\rho\colon X\times\X\to[0,1]$ be a stochastic choice on a set of cardinality $\vert X\vert\geq 3$. The following are equivalent:
	\begin{enumerate}[\rm(i)]
		\item $(\rhd,Pr_{\rho,\rhd})$ is the unique justification by compromise of $\rho$;
		\item $\rhd$ composes $\rho$, and one of the following conditions hold:
	 \begin{itemize}
	 	\item[(a)] $\vert X^{*} \vert \geq 3$;
	 	\item[(b)] $\vert  X^{*} \vert=2$, and $\min(X,\rhd)\not\in X^{*}.$\footnote{Davide Carpentiere provided some results that dramatically shortened the proof of this theorem.}
	 	 		 	\end{itemize}
	 
	 \end{enumerate}
\end{theorem}

Theorem~\ref{THM:stochastic_self_punishment_identification} states that a stochastic choice $\rho$ on $X$ has a unique justification by compromise if and only if her true preference $\rhd$ composes $\rho$, and there are at least three items that are selected from $X$ with non-zero probability, or only two alternatives, both distinct from $\min(X,\rhd),$  are chosen with positive probability from $X$.
This result allows to retrieve from data, in all but two cases, the endogenous parameters of my model, i.e., the DM's taste, and her randomization among its compromises.
Moreover, when a linear order $\rhd$ composes $\rho$,  and only $\min(X,\rhd)$ and another item  are selected from $X$ with non-zero probability, there is only another distinct justification by compromise of the dataset, in which the other DM's underlying preference can be derived from $\rhd$.
To see this, I need some additional notation.
Given a linear order $\rhd\in\mathsf{LO}(X)$, and some $j\in\{1,\cdots,\vert X\vert\}$, denote by $\rhd^{*j}$ the linear order defined by $x^{\rhd^{j*}}_h=x^{\rhd}_{h},$ for all $1\leq h <j$, and  $x^{\rhd^{*j}}_h=x^{\rhd}_{\vert X\vert-h+j}$ for any $  j\leq h\leq \vert X\vert.$
The preference $\rhd^{*j}$ is generated from $\rhd$ by keeping fixed the ranking of the first $j-1$ items, and inverting the ranking of the other $\vert 	X\vert-j+1$ alternatives.
We have:

\begin{lemma}\label{LEMMA:unique_two_justifications_two_items}
Assume that $\rhd\in\mathsf{LO}(X)$ composes $\rho\colon X\times\X\to[0,1] ,$ $\vert  X^{*} \vert=2$, and $\min(X,\rhd)\in X^{*}.$
Let $j\in\{1,\cdots,\vert X\vert-1\}$ be the other index such that $\rho(x^{\rhd}_{j},X)>0.$
Then  $(\rhd, Pr_{\rho,\rhd})$ and $(\rhd^{*j}, Pr_{\rho,\rhd^{*j}})$ are the only two justifications by compromise of $\rho.$
Moreover, I have that $Pr_{\rho,\rhd}(\rhd_{j-1})=Pr_{\rho,\rhd^{*j}}\left(\rhd^{*j}_{\vert X\vert-1}\right)> 0,$ and $Pr_{\rho,\rhd}(\rhd_{\vert X\vert-1})=Pr_{\rho,\rhd^{*j}}\left(\rhd^{*j}_{j-1}\right)> 0.$   
\end{lemma}

{Lemma~\ref{LEMMA:unique_two_justifications_two_items} applies also to the compromise-based RUMs whose support contains only two opposite criteria.
However, this limitation is consistent with the behavioral interpretation of the model: when the DM wavers only between two conflicting views,  she may be not aware of which of the two is her \textit{true} preference.}

If a linear order $\rhd$ composes $\rho$, and there is only {an} item which is selected with positive probability from $X$, then identification vanishes, and the dataset has at least $\vert X\vert$ distinct justifications by compromise.

\begin{lemma}\label{LEMMA:unique_two_justifications_one_item}
Assume that $\rhd\in\mathsf{LO}(X)$ composes $\rho\colon X\times\X\to[0,1] ,$ and $\vert  X^{*} \vert=1$.
Then for any $j\in\{0,\cdots,\vert X\vert-1\}$ there is $\rhd^{\prime}\in\mathsf{LO}(X)$ such $(\rhd^{\prime},Pr_{\rho,\rhd^{\prime}})$ justifies by compromise $\rho$, and $Pr(\rhd^{\prime}_j)=1.$  
\end{lemma}

The above findings indicate that for most of the compromise-based RUMs the experimenter can unambiguously pin down the DM's true preference and the compromises adopted in the decision, and observe the extent of her mediation.
In the next subsection I propose a measure of the DM's mediation needed to explain stochastic choice data.
Theorem~\ref{THM:stochastic_self_punishment_identification}, Lemma~\ref{LEMMA:unique_two_justifications_two_items}, and Lemma~\ref{LEMMA:unique_two_justifications_one_item} are crucial to reduce the computational complexity of this test. 

\subsection{Degree of compromise}\label{SUBSECTION:degree_self_punishment}

{If the observed choice can be explained by appealing to compromise, the experimenter could be interested into estimating the extent of the DM's mediation.
To do so, I propose a measure of compromise, consisting of the maximum number of alternatives on top of the DM's true preference which have been disregarded to perform her selection. 

\begin{definition}\label{DEF:stochastic_degree_self_punishment}
	 Given a compromise-based RUM $\rho\colon X\times \X\to [0,1]$, I denote by
	$$cp(\rho)=\min_{(\rhd,Pr)\in\mathsf{JC}_{\rho}}\left(\max_{i\colon Pr(\rhd_i)>0} i\right)$$
	the \textsl{degree of compromise} of $\rho$.
\end{definition}

The degree of compromise is the minimum value, among all the pairs $(\rhd,Pr)$ that justify by compromise $\rho$, of the maximal index $i$ of the compromises that have been selected with positive probability.
It estimates a lower bound to the maximal level of mediation that the DM has adopted in her decision.
The computation of $cp$ relies on the following property.

\begin{definition}\label{DEF:ordered_composition_degree_j}
	A stochastic choice $\rho\colon X\times \X\to[0,1] $ \textit{has a $j$-th ordered composition} if there is some linear order $\rhd\in\mathsf{LO}(X)$ that composes $\rho$, $\rho\left(x^{\rhd}_j,X\right)\neq 0$ for some $1\leq j\leq \vert X\vert$, and $\rho\left(x^{\rhd}_{l},X\right)=0$ for any $ j<l\leq \vert X\vert$.
\end{definition} 

Thus, $\rho$ has a $j$-th ordered composition  if there is a linear order $\rhd$ that composes $\rho$ such that $x^{\rhd}_j$ is selected with positive probability, and the probability of selecting any item worse than $x^{\rhd}_{j}$ from the ground set is null.
It is evident that if a compromise-based RUM on a set $X$ has a degree of compromise equal to $i$, then it has a $(i+1)$-th ordered composition.
Remarkably, the inverse implication is also true, if there are at least two items which have been selected with non-zero probability from the ground set.

\begin{theorem}
	
\label{THM:computation_stochastic_degree_self_punishment}

	Let $\rho\colon X\times \X\to [0,1]$ be a compromise-based RUM defined on a ground set of cardinality $\vert X \vert\geq 3$.
	If $\vert X^{*}\vert=1,$ then $cp(\rho)=0$.
If $\vert X^{*}\vert\geq 2,$ then, given $0\leq i\leq \vert X\vert-1$, I have that $cp(\rho)=i$ if and only if $\rho$ has a $(i+1)$-th ordered composition.
	\end{theorem}

Theorem~\ref{THM:computation_stochastic_degree_self_punishment} shows how to elicit from data the maximum level of mediation that DM applied for sure in her decision. 
When there are at least two items that have been selected with non-zero probability from the ground set, this measure captures exactly the extent of the DM's compromise.

\begin{lemma}\label{LEMMA:degree_self_punishment_exact}
Let $\rho\colon X\times \X\to [0,1]$ be a compromise-based RUM defined on a ground set of cardinality $\vert X \vert\geq 3$.
If $\vert X^{*}\vert\geq 2,$ then 
 		$$cp(\rho)=\max_{i\colon Pr(\rhd_i)>0} i$$
 	for any $(\rhd,Pr)\in\mathsf{JC}_{\rho}$.	
\end{lemma}

In this case, the computation of the degree of compromise comes after the identification.
The experimenter first derives, by implementing the compromise-based revealed preference algorithm, a linear order that composes the dataset.
Then, he observes the $i$ for which, given the ranking he found, the analyzed stochastic choice has a $(i+1)$-th ordered composition, and he can deduce that $i$ is  the degree of compromise.} 

\begin{remark}
{The index presented in Definition~\ref{DEF:stochastic_degree_self_punishment} throws away some cardinal information, since it does not take into account choice frequencies.
An alternative measure of mediation between opposite judgments may be the \textit{frequentist degree of compromise}, denoted by $f\!cp$ and defined by  
$$f\!cp(\rho)=\min_{(\rhd,Pr)\in\mathsf{JC}_{\rho}}\sum_{i=0}^{\vert X\vert-1}Pr(\rhd_i)\,i$$
for any compromise-based RUM $\rho\colon X\times \X\to [0,1]$.\footnote{I thank an anonymous referee for proposing this statistic.}
Instead of disclosing the maximal extent of DM's mediation, the frequentist degree of compromise gauges its \textit{intensity}, by weighting all the indices of the compromises justifying her choice for the probabilities with which these linear orders are adopted.
This score can be easily computed, once the experimenter has elicited the rankings composing the dataset, by using the algorithm of Definition~\ref{DEF:revealed_preference_algorithm_self_punishment}, and inferred, applying Corollary~\ref{COR:derivation_probability_from_dataset}, the probability distributions over compromises.
Moreover, Theorem~\ref{THM:stochastic_self_punishment_identification} ensures that, when $\vert X^*\vert \geq 3$, $f\!cp$ specifies precisely the magnitude of the DM's mediation.}  	
\end{remark}

In the next subsection  I explore the connections between compromise-based RUMs and other subclasses of RUMs that have been discussed in the literature.   

\section{Relation with the literature}\label{SECT:relation_literature}
Compromise-based RUMs are RUMs whose support is limited to the compromises on some preference.
However, not all RUMs are compromise-based, as showed in the next example.

\begin{example}\label{EXMP:RUM_functions_not_compromise-based}
Let $X=\{x,y,z\}$ and $\rho\colon X\times\X\to[0,1]$ be the stochastic choice defined as follows:
	\medskip
	\begin{center}
\begin{tabular}{ ccccccccccccc} 
\hline
   & $X$  & $xy$ & $xz$ & $yz$ \\ 
\hline
$x$ & $0.2$ & $0.6$ & $0.2$ & $0$\\ 
\hline
$y$ & $0.2$ & $0.4$ & $0$ & $0.4$\\
\hline
$z$ &$0.6$ & $0$ & $0.8$ & $0.6$\\
\hline
\end{tabular}
\end{center}
\medskip	

The above dataset is not {justified by compromise}, since the procedure described in Definition~\ref{DEF:revealed_preference_algorithm_self_punishment} cannot be even started.
Indeed, there is no item whose probability of selection is constant across all menus of cardinality greater than one containing it.
However, $\rho$ is a RUM.
Given the linear orders $\rhd\colon x\rhd y\rhd z$, and $\rhd^{\prime}\colon z\rhd^{\prime} x \rhd^{\prime} y$,
 the probability distribution $Pr \in\Delta(\mathsf{LO}(X))$ with support $Pr(\rhd)=Pr(\rhd_1)=Pr(\rhd_2)=0.2$, and $Pr(\rhd^{\prime})=0.4$, rationalizes $\rho$. 
\end{example}    

\cite{ApesteguiaBallesterLu2017} analyze RUMs whose support is a collection of preferences satisfying the \textit{single crossing property}.
More formally, given a set $X$ linearly ordered by $\rhd\in\mathsf{LO}(X)$, a stochastic choice $\rho\colon X\times\X\to[0,1]$ is a \textit{single crossing RUM} if it is a RUM, and it is explained by some $Pr\in\Delta(\mathsf{LO}(X))$, whose support can be ordered as $\left(\rhd^1,\cdots,\rhd^{T}\right)$ to satisfy the following condition: for any $s,t\in\{1,\cdots,T\}$ such that $s<t$, and for any $x,y\in X$ such that $x\rhd y$, if $x\rhd^s y$, then $x\rhd^t y$.
The authors also investigate RUMs explained only by \textit{single peaked preferences}.
Given a set $X$ linearly ordered by $\rhd\in\mathsf{LO}(X)$, a stochastic choice  $\rho\colon X\times\X\to[0,1]$ is a \textit{single peaked RUM} if it is a RUM, and it is explained by some $Pr\in\Delta(\mathsf{LO}(X))$ such that every $\rhd^{\prime}$ for which $Pr(\rhd^{\prime})>0$ is \textit{single peaked with respect to} $\rhd$, i.e., $y\rhd x\rhd \max(X,\rhd^{\prime})$ or $\max(X,\rhd^{\prime})\rhd x\rhd y$ implies $x\rhd^{\prime} y$.
The class of single peaked RUMs is a subclass of single crossing RUMs.
As expected, any stochastic choice $\rho\colon X\times\X\to[0,1]$ that is justified by compromise by some pair $(\rhd,Pr)$ is a single peaked RUM, and, thus, it is a single crossing RUM.
To see why, note that if I assume that $X$ is linearly ordered by $\rhd$, by Definition~\ref{DEF:compromise} I have that, for any $\rhd_i\in\mathsf{Comp}(\rhd)$, if $y\,\rhd x\rhd\max(X,\rhd_i)$, then $x\rhd_i y$. 
The same happens if $\max(X,\rhd_i)\rhd x \rhd y$.
However, there are single peaked RUMs that are not compromise-based RUMs, as showed in the following example.

\begin{example}
	Let $X=\{w,x,y,z\}$ and $\rho\colon X\times\X\to[0,1]$ be the stochastic choice defined as follows:
	\medskip
	\begin{center}
\begin{tabular}{ ccccccccccccc} 
\hline
   & $X$  & $wxy$ & $wyz$ & $wxz$ & $xyz$ & $wx$ & $wy$ & $wz$ & $xy$ & $xz$ & $yz$ \\ 
\hline
$w$ & $0.8$& $1$& $0.8$& $0.8$& $0$& $1$& $1$& $0.8$& $0$& $0$& $0$  \\ 
\hline
$x$ & $0$ & $0$ &$0$ & $0$ &  $0$ & $0$ & $0$ & $0$ & $0$ &$0.6$ & $0$\\ 
\hline
$y$ & $0$ & $0$ & $0$ & $0$ & $0.6$ & $0$ & $0$ &$0$ & $1$ & $0$ & $0.6$\\
\hline
$z$ &$0.2$ & $0$ & $0.2$ & $0.2$ & $0.4$ & $0$ & $0$ & $0.2$ & $0$ & $0.4$ & $0.4$ \\
\hline
\end{tabular}
\end{center}
\medskip	

The dataset $\rho$ is a single peaked RUM.
To see why, let $\{\rhd,\rhd^{\prime},\rhd^{\prime\prime}\}$ be a collection of linear orders defined by $\rhd\colon z\,\rhd\, w\rhd\, y\rhd\, x,$ $\rhd^{\prime}\colon w\,\rhd^{\prime} \,z\rhd^{\prime} \,y\rhd^{\prime}\,x,$ $\rhd^{\prime\prime}\colon w\rhd^{\prime\prime}y\rhd^{\prime\prime} x\rhd^{\prime\prime} z$, and let $Pr\in\Delta(\mathsf{LO}(X))$ be such that $Pr(\rhd)=0.2, Pr(\rhd^{\prime})=0.2, Pr(\rhd^{\prime\prime})=0.6.$
One can check that $\rho$ is a RUM, $Pr$ rationalizes $\rho$, and, considered the set $X$ linearly ordered by $\rhd$, each linear order of the collection $\{\rhd,\rhd^{\prime},\rhd^{\prime\prime}\}$ if single peaked with respect to $\rhd$.
However, $\rho$ is not a compromise-based RUM, because the compromise-based revealed preference algorithm cannot be completed (again, there is no item in $X$ whose probability of selection is constant among all menus $A\in\X$ containing it and having size $\vert A\vert\geq 2$).
Remarkably, and differently from single crossing and single peaked RUMs, in my model the DM's preference is an \textit{endogenous} parameter, that, as showed by Theorem~\ref{THM:stochastic_self_punishment_identification}, can be retrieved from data.   
\end{example} 

\cite{MariottiManzini2018} and \cite{MariottiManziniPetri2019} discuss \textit{menu-independent dual RUMs}, i.e. RUMs rationalized by two linear orders.\footnote{The authors also investigate \textit{menu-dependent dual RUMs}, in which the randomization over the two linear orders may change across menus.
Since menu-dependent dual RUMs are not a proper subclass of RUMs, I do not include them in this survey.}
Compromise-based RUMs and dual RUMs are independent families of stochastic choices.
As a matter of fact, there are compromise-based RUMs that are not menu-independent dual RUMs.
For instance, the compromise-based RUM displayed in Example~\ref{EXMP:temptation} is a uniquely identified RUM, rationalized by a probability distribution that assumes positive values only on three distinct linear orders.
Thus, it is not a menu-independent dual RUM.
Moreover, there are menu-independent dual RUMs that are not compromise-based.
To see this, consider some RUM $\rho\colon X\times\X\to[0,1]$ rationalized by a probability distribution $Pr\in\Delta(\mathsf{LO}(X))$ which assumes positive values only on the linear orders $\rhd^{\prime},\rhd^{\prime\prime}\in\mathsf{LO}(X)$, respectively defined by $x\rhd^{\prime} y\rhd^{\prime} z,$ and $x\rhd^{\prime\prime} z\rhd^{\prime\prime} y.$
By definition, $\rho$ is a menu-independent dual RUM.
The reader can check that there is no $\rhd\in\mathsf{LO}(X)$ such that $\{\rhd^{\prime},\rhd^{\prime\prime}\}\subset \mathsf{Comp}(\rhd)$, thus $\rho$ is not a compromise-based RUM explained only by two compromises on some preference.
Since compromise-based RUMs are uniquely identified RUMs, I conclude that $\rho$ is not a compromise-based RUM.   
 
\cite{Turansick2022} offers two characterizations of uniquely identified RUMs.  
In the proof of Lemma~\ref{LEM:uniqueness_up_to_probability_distribution_preliminary} I use one of his results to prove that any compromise-based RUM is a uniquely identified RUM.
However, there are uniquely identified RUMs that are not compromise-based.
Indeed, \cite{BlockMarschak1960} and \cite{Turansick2022} show that any RUM on a ground set of size $\vert X\vert\leq 3$ is uniquely identified.
Thus, the RUM displayed in Example~\ref{EXMP:RUM_functions_not_compromise-based} is rationalized by a unique probability distribution, but it is not compromise-based.\footnote{\cite{Turansick2022} exhibits also some single-crossing RUMs that are not uniquely identified RUMs.}

\cite{Valkanova2024} introduces four subclasses of RUMs, respectively called \textit{peak-pit on a line}, \textit{locally peak-pit}, \textit{triple-wise value-restricted}, and \textit{peak-monotone RUMs}. 
A stochastic choice $\rho\colon X\times\X\to[0,1]$ is a \textit{peak-pit on a line RUM} if it is a RUM, and there is a $Pr\in\Delta(\mathsf{LO}(X))$ that rationalizes it, and a linear order $\rhd\in \mathsf{LO}(X) $ such that, for every $\rhd^{\prime}$ for which $Pr(\rhd^{\prime})>0$, and any $\{x,y,z\}\subseteq X$ for which $z=\max(xyz,\rhd^{\prime})$, $x\rhd y\rhd z$ or $z\rhd y\rhd x$ implies $y\rhd^{\prime} x$, provided that there is $\rhd^{\prime\prime}\in \mathsf{LO}(X)$ such that $Pr(\rhd^{\prime\prime})>0$, and $y=\max(xyz,\rhd^{\prime\prime})$.
Moreover, $\rho\colon X\times\X\to[0,1]$ is a \textit{locally peak-pit RUM} if it is a RUM, and there is a $Pr\in\Delta(\mathsf{LO}(X))$ that rationalizes it such that, for every $\{x,y,z\}\subseteq X$ and some $x^*\in \{x,y,z\}$, there is no $\rhd\in \mathsf{LO}(X)$ for which $Pr(\rhd)>0$, and $x^*=\max(xyz,\rhd)$, or there is no $\rhd\in \mathsf{LO}(X)$ for which $Pr(\rhd)>0$, and $x^*=\min(xyz,\rhd)$.
A stochastic choice $\rho\colon X\times\X\to[0,1]$ is a \textit{triple-wise value-restricted RUM} if it is a RUM, and there is a $Pr\in\Delta(\mathsf{LO}(X))$ that rationalizes it such that, for every $\{x,y,z\}\subseteq X$ and some $x^*\in \{x,y,z\}$, there is no $\rhd\in \mathsf{LO}(X)$ for which $Pr(\rhd)>0$, and $x^*=\max(xyz,\rhd)$, or there is no $\rhd\in \mathsf{LO}(X)$ for which $Pr(\rhd)>0$, and $x^*=\min(xyz,\rhd)$, or there is no $\rhd\in \mathsf{LO}(X)$ for which $Pr(\rhd)>0$, $x^*\neq\max(xyz,\rhd)$, and $x^*\neq\min(xyz,\rhd)$.
Finally, $\rho\colon X\times\X\to[0,1]$ is a \textit{peak-monotone RUM} if it is a RUM, and there are a $Pr\in\Delta(\mathsf{LO}(X))$ that rationalizes it, and a linear order $\rhd\in\mathsf{LO}(X)$ such that, for every $\rhd^{\prime},\rhd^{\prime\prime}\in\mathsf{LO}(X)$ for which $Pr(\rhd^{\prime})>0$ and $Pr(\rhd^{\prime\prime})>0$, and any  $\{x,y,z\}\subseteq X$ with $z=\max(xyz,\rhd^{\prime})$, and $z=\max(X,\rhd^{\prime\prime}),$ I have that 

$$x\rhd y\rhd z\;\text{or}\; z\rhd y\rhd x \;\text{implies}\; y\rhd^{\prime} x,$$
provided that there are $\rhd^{\prime\prime\prime},\rhd^{\prime\prime\prime\prime}\in\mathsf{LO}(X)$ for which $Pr(\rhd^{\prime\prime\prime})>0$ and $Pr(\rhd^{\prime\prime\prime\prime})>0$ such that $y=\max(X,\rhd^{\prime\prime\prime})$ and $x=\max(X,\rhd^{\prime\prime\prime\prime}),$ and

$$x\rhd y\rhd z\;\text{or}\; z\rhd y\rhd x \;\text{implies}\; y\rhd^{\prime\prime} x,$$
provided that there is $\rhd^{\prime\prime\prime}\in\mathsf{LO}(X)$ for which $Pr(\rhd^{\prime\prime\prime})>0$ such that $y=\max(X,\rhd^{\prime\prime\prime})$.
The author shows that single peaked RUMs are peak-pit on a line RUMs, which, in turn, are locally peak-pit RUMs, triple-wise value-restricted RUMs, and peak-monotone RUMs.\footnote{Moreover, locally peak-pit RUMs are triple-wise value-restricted RUMs, and these two subclasses are non-nested with peak-monotone RUMs.}
It follows that compromise-based RUMs are a subclass of these four specifications.

\cite{CaliariPetri2024} investigate special RUMs, called \textit{irrational RUMs}, which are generated by probability distributions over deterministic choice functions that violate WARP.
The authors show that each stochastic choice $\rho\colon X\times\X\to[0,1]$ is an irrational RUM if and only if \textit{Correlation Bounds} is satisfied, i.e., denoted by $\X(\rhd)$ the family of menus $\left\{A\in\X\colon \vert A\vert\geq 2\right\}\setminus \{\min(X,\rhd),\max(X,\rhd)\},$ the condition 
$$\mathbb{C}^{\,\rho}_{\,\rhd}=\frac{1}{\vert \X(\rhd)\vert-1 }\sum_{A\in \X(\rhd)}\rho(\max(A,\rhd), A)\leq 1$$
holds for any $\rhd\in\mathsf{LO}(X)$.
Irrational RUMs and compromise-based RUMs are non-nested subclasses of RUMs.
Indeed, some irrational RUMs are not compromise-based RUMs. 
As an illustration of this, note that the dataset displayed in Example~\ref{EXMP:RUM_functions_not_compromise-based} satisfies Correlation Bounds, but it is not compromise-based.
Moreover, there are compromise-based RUMs that are not irrational RUMs, as showed in the next example.
\begin{example}\label{EXMP:self_punishment_irrational_}
Let $X=\{x,y,z\}$ and $\rho\colon X\times\X\to[0,1]$ be the stochastic choice defined by
	\medskip
	\begin{center}
\begin{tabular}{ ccccccccccccc} 
\hline
  & $X$  & $xy$ & $xz$ & $yz$ \\ 
\hline
$x$ & $0.95$ & $0.95$ & $0.95$ & $0$\\ 
\hline
$y$ & $0.05$ & $0.05$ & $0$ & $1$\\
\hline
$z$ &$0$ & $0$ & $0.05$ & $0$\\
\hline
\end{tabular}
\end{center}
	
The dataset $\rho$ is a compromise-based RUM, and it is {justified by compromise} by the pair $(\rhd,Pr)$, with $\rhd\colon x\rhd y\rhd z$, $Pr(\rhd)=0.95$, and $Pr(\rhd_1)=0.05$.
We also have that $\rho$ is not an irrational RUM, since $\mathbb{C}^{\,\rho}_{\,\rhd}=1.45>1$.
\end{example}

\cite{Suleymanov2024} discusses a subclass of RUMs that have a \textit{branching independent RUM representation}, i.e., for every preference of the support of the probability distribution that rationalizes the dataset, and fixed the first $k$ and the last $n- k$ items, the relative ordering of the first $k$ elements is independent of the relative ordering of the last $n- k$ elements.
More formally, given a linear order $\rhd\in\mathsf{LO}(X)$, I denote by $P_k^{\,\rhd}$ and $D_{k}^{\,\rhd}$ respectively the first $k$ and the
last $\vert X\vert-k+1$ ranked alternatives according to $\rhd$.
Given a set $A\in\X$, I denote by $\rhd^{\downarrow}_A$ the restriction of $\rhd$ to $A$.
Moreover, $\rhd^{\prime}$ is a \textit{$k$-branching} of $\rhd$ if $P_k^{\,\rhd}=P_k^{\,\rhd^{\prime}}$ holds, and I denote by $B^{\,\rhd}_k$ all the $k$- branching of $\rhd$.
  A probability distribution $Pr\in\Delta(\mathsf{LO}(X))$ is \textit{branching independent} if for any $\rhd\in\mathsf{LO}(X)$ such that $Pr(\rhd)>0$ and
 $1\leq k\leq \vert X\vert -1$ I have that 
 
 {\small\begin{equation}\label{EQ:branching_independent_distribution}
 Pr\left(\rhd^{\prime}=\rhd\Big\vert \,\rhd^{\prime}\in B_k^{\,\rhd}\right)=Pr\left({\rhd^{\prime\,\,}}^{\downarrow}_{P_{k}^{\,\rhd}}=\rhd^{\downarrow}_{P_{k}^{\,\rhd}}\Big\vert\, \rhd^{\prime} \in B^{\,\rhd}_k\right)\cdot Pr\left({\rhd^{\prime\,\,}}^{\downarrow}_{D_{k+1}^{\,\rhd}}=\rhd^{\downarrow}_{D_{k+1}^{\,\rhd}}\Big\vert \,\rhd^{\prime} \in B^{\,\rhd}_k\right).
  \end{equation}}
  
  Then $\rho\colon X\times\X\to[0,1]$ has a branching independent RUM representation if there is a branching independent probability distribution $Pr\in\mathsf{LO}(X)$ that rationalizes $\rho$.
The author proves that any RUM is a stochastic choice having a branching independent RUM representation, and vice versa.
Moreover for each RUM, the branching independent RUM representation is unique. 
Since compromise is nested in RUMs, it is also nested in the class of stochastic choices with branching independent RUM representation.
The connection between the framework \cite{Suleymanov2024} and compromise-based RUMs is clarified by the following insight: given a linear order $\rhd\in\mathsf{LO}(X)$, note that for each $0\leq i\leq \vert X\vert-1$ I have that 
\begin{equation}\label{EQ:branchings_of_a_compromise-based_distortion}
  B^{\,\rhd_i}_k =
    \begin{cases}
      \{\rhd_h\colon 0< k\leq h \} & \text{if $\vert X\vert-i-1 \leq  k\leq\vert X\vert-1$,}\\
      \{\rhd_i\} & \text{if $1\leq k<\vert X\vert-i-1$.}
    \end{cases}       
\end{equation}

Assume now that $\rho\colon X\times\X\to[0,1]$ is compromise-based, and that the pair $(\rhd,Pr)$ explains $\rho$ by compromise.
Thus, for each $i,j\in\{0,\cdots,\vert X\vert-1\}$ such that $Pr(\rhd_i)>0$, and any $1\leq k\leq \vert X\vert -1$, Equality~(\ref{EQ:branching_independent_distribution}) can be rewritten as
  {\small \begin{equation*}
 	Pr\left(\rhd_j=\rhd_i\Big\vert \,\rhd_j\in B_k^{\,\rhd_i}\right)=Pr\left({\rhd_j}^{\downarrow}_{P_{k}^{\,\rhd_i}}={\rhd_i}^{\downarrow}_{P_{k}^{\,\rhd_i}}\Big\vert\, \rhd_j\in B^{\,\rhd_i}_k\right)\cdot Pr\left({\rhd_j}^{\downarrow}_{D_{k+1}^{\,\rhd_i}}={\rhd_i}^{\downarrow}_{D_{k+1}^{\,\rhd_{i}}}\Big\vert \,\rhd_j \in B^{\,\rhd_i}_k\right),
 	 \end{equation*}}
 which, by Equality~(\ref{EQ:branchings_of_a_compromise-based_distortion}) and Definition~\ref{DEF:compromise}, gives 
$$\begin{cases}
    \frac{Pr(\rhd_i)}{\Sigma_{k\leq h}Pr(\rhd_h)}=\frac{Pr(\rhd_i)}{\Sigma_{k\leq h}Pr(\rhd_h)}\cdot \frac{Pr(\rhd_i)}{Pr(\rhd_i)}=\frac{Pr(\rhd_i)}{\Sigma_{k\leq h}Pr(\rhd_h)} & \text{if $\vert X\vert-i-1 \leq  k\leq\vert X\vert-1$,}\\
      \frac{Pr(\rhd_i)}{Pr(\rhd_i)} = \frac{Pr(\rhd_i)}{Pr(\rhd_i)} \cdot \frac{Pr(\rhd_i)}{Pr(\rhd_i)} & \text{if $1\leq k<\vert X\vert-i-1$.}
    \end{cases}$$ 
 
 The comparison with stochastic choices having a branching independent RUM representation concludes this section, whose main findings are summarized in the following diagrams.   
    
\vspace{1.5cm}
\begin{figure}[H]
\begin{pspicture}(-3.5,-2,36)(0.48,3)
\pscircle[linewidth=1pt, linecolor=black](2.03,1.2){3.8}
\pscircle[linewidth=1pt, linecolor=black](2.03,1.2){2.2}
\pscircle[linewidth=1pt, linecolor=black](2.03,1.2){1.5}

  % Intersezione A e C
  \psellipse[linewidth=1pt, gradangle=45,  gradbegin=blue!20, gradend=white, 
           , linewidth=1pt](0.86,1.2)(2.5,0.6)
   \psellipse[linewidth=1pt, gradbegin=red!20, gradend=white, 
            linecolor=red
           , linewidth=1pt](2,1.19)(0.8,0.35)
   \psellipse[linewidth=1pt, rot=-30, gradangle=45,  gradbegin=blue!20, gradend=white, 
           , linewidth=1pt](2.4,1.9)(1.8,2.5)
 
 \psellipse[linewidth=1pt, gradangle=45,  gradbegin=red!20, gradend=white, linecolor=red
           , linewidth=1pt](9.5,1.19)(1,0.35)
    \psellipse[linewidth=1pt, gradangle=45,  gradbegin=blue!20, gradend=white, 
           , linewidth=1pt](7.9,1.2)(1.3,0.4)
            \psellipse[linewidth=1pt, rot=25, gradangle=45,  gradbegin=blue!20, gradend=white, 
           , linewidth=1pt](8.1,0.5)(1.3,0.4)

  % Etichette degli insiemi
\rput(2,4.5){\footnotesize RUMs $\equiv$ biRUMs }
\rput(2.5,3.7){\footnotesize pmRUMs}
%\rput(2,4.35){\footnotesize }
 %\rput(2,3.7){\footnotesize LPPRUMs}
 \rput(2,2.95){\footnotesize twvrRUMs}
 \rput(2,2.2){\footnotesize scRUMs}
 \rput(-0.8,1.2){\footnotesize uiRUMS}
 %\rput(-2.87,0.33){\footnotesize idRUMs}
 \rput(2,1.22){\footnotesize cbRUMs}
  \rput(9.8,1.22){\footnotesize\,cbRUMs}
   \rput(7.7,1.22){\footnotesize midRUMs}
    \rput(7.8,0.36){\footnotesize iRUMs}
\end{pspicture}
\caption{Compromise-based RUMs (cbRUMs) are a subclass of single-crossing RUMs (scRUMs), triple-wise value-restricted RUMs (twvrRUMs), peak-monotone RUMs (pmRUMs), uniquely identified RUMs (uiRUMs), and stochastic choices with a branching independent RUM representation (biRUMs).
The subclass of cbRUMs is independent of irrational RUMs (iRUMs) and menu-independent dual RUMs (midRUMs).}
\end{figure}
 \label{FIG:relation_models}        

\section{Concluding remarks}\label{SECT:concluding_remarks}

{In this paper I assume that a mediation between opposite views settles individual choice, and induces the DM to apply some compromises on her true preference, evaluating some of the best alternatives according to the antithetical criterion.
Compromise-based RUMs, which are RUMs whose support is limited to the compromises on some preference, are characterized by the existence of a linear order that allows to recover choice probabilities from the DM's selection over the ground set.
Given this ranking, some properties of the dataset, which are related to the phenomenon examined, are necessary and sufficient.
I provide a simple test for compromise-based RUMs, which also elicits the DM's preferences supporting data.}
Compromise-based RUMs are uniquely identified RUMs.
However, {a unique justification by compromise} is admitted if and only if there is a linear order the composes data, and the DM's selects with non-zero probability from the ground set either at least three items, or only two alternatives, both distinct from the minimal option.
{I define a degree of compromise, estimating the scope of the DM's mediation, and I propose a characterization of it.}
Finally, the relationships between compromise-based RUMs are other subclasses of RUMs are examined.   

{In my framework the DM's mediation between conflicting judgements is random, and there is no rule that matches menus and the maximizing compromises.}
Thus, future research should be devoted to describe the \textit{causes} of this trade-off, by formally defining  a mechanism that associates compromises to menus.
{Moreover, one can imagine a \textit{menu-dependent} mediation.
 Indeed, the DM could apply a \textit{random attention rule} \citep{Cattaneoetal2020} that from each menu discards, with a probability satisfying some consistency condition, the first $i$ items on top of her preference, avoiding overly  neglecting the opposite criterion.
It would be interesting to compare the explanatory power and the welfare indications of this suggested approach with compromise-based RUMs.   
A menu-dependent model of compromise may, for instance, disclose the DM's tendency to exclude from her choice set \textit{tempting alternatives}, as in \cite{GulPesendorfer2001,GulPesendorfer2004}, \cite{DekelLipmanRustichini2009}, and \cite{Noor2011}.} 
{Finally, a potential extension of my setting may account for \textit{dynamic} compromise, in which the DM's propensity to adopt compromises on her preference in a given period depends also on the 
mediation experienced in the past.
This pattern could explain \textit{intertemporal stochastic choices}, recently discussed by \cite{FudengbergIijimaStrzalecki2015} and \cite{LuSaito2018}.}

{\section*{Appendix: Proofs}
\noindent \textbf{\large Proof of Lemma~\ref{LEM:utility_representation_compromise_based_RUMs}}.
I need some preliminary notation.   
Order the ground set $X$ as $\left\{x^{\rhd}_1,\cdots,x^{\rhd}_{\vert X\vert}\right\},$ where $x^{\rhd}_i\rhd x^{\rhd}_j$ if and only if $i<j.$
Thus, given some $1\leq j\leq \vert X\vert$, $x_j^{\rhd}$ denotes the $j$-th item of $X$ with respect to $\rhd$.
Moreover, denote by $x_j^{\uparrow\rhd}$ the set $\{y\in X \colon y\,\rhd x^{\rhd}_j \}=\{x^{\rhd}_h\in X\colon h< j\}$, by $x_j^{\downarrow\rhd}$ the set $\{y\in X \colon x^{\rhd}_j\rhd y \}=\{x^{\rhd}_k\in X\colon k> j\}.$

(i)$(\Longrightarrow)$(ii).
	Define $U(x)=\vert x^{\downarrow\rhd} \cup x\vert $ for any $x\in X$. 
Note that since $U$ is always positive, we have that $-U(y)<U(z)$ for any $y,z\in X$.
Moreover, since $\rhd$ is a linear order on $X$, $U$ is injective, and $U(y)>U(z)$ if and only if $y\,\rhd z$ for any $y,z\in X.$
For any $i\in\{0,\cdots,\vert X\vert-1\}$ such that $\rhd_i\in\mathsf{Comp}(\rhd)$ and $Pr(\rhd_i)>0$, set $\alpha_i=\vert {x_i}^{\downarrow\rhd} \cup x^{\rhd}_i \vert=U(x^{\rhd}_i)$.
Denote by $\mathcal{T}$ the set $\{\alpha_i\}_{i\colon Pr(\rhd_i)>0}$, and  by $P\in\Delta(\mathcal{T})$ the probability distribution defined by $P(\alpha_i)=Pr(\rhd_i)$ for any $i\in\{0\,\cdots,\vert X\vert-1\}$ such that $Pr(\rhd_i)>0$.\footnote{Since $Pr\in\Delta(\mathsf{Comp}(\rhd))$ is a probability distribution, $P\in\Delta(\mathcal{T})$ is a probability distribution.}
Note that $\vert \mathcal{T}\vert\leq \vert X\vert$.
Consider now $C_{\alpha_i}\colon X\to \mathbb{R}$ defined by  $C_{\alpha_i}(x)=U(x)\mathbf{1}_{\left\{U(x)< \alpha_i \right\}}-U(x)\mathbf{1}_{\left\{U(x)\geq \alpha_i \right\}}$ for any $x\in X.$
We show that, for any $i\in\{0,\cdots,\vert X\vert-1\}$, and any $x,y\in X$, we have that $x\rhd_i y$ if and only if $C_{\alpha_i}(x)>C_{\alpha_i}(y)$.

Assume that $x\rhd_i y$ for some $i\in\{0,\cdots,\vert X\vert-1\}$ and $x,y\in X$.
By Definition \ref{DEF:compromise}, three mutually exclusive cases are possible: 1) $x,y\in X^{\rhd}_i$ and $y\rhd x$, 2) $y\in X^{\rhd}_i$, $x\not\in X^{\rhd}_i$, and $y\rhd x$, or $x,y\not\in X^{\rhd}_i$ and $x\rhd y$.
If 1) holds, then the definition of $U$ yields $U(y)>U(x)\geq U(x^{\rhd}_i)=\alpha_i$.
The definition of $C_{\alpha_i}(x)$ and the fact that $U$ is always positive imply that $C_{\alpha_i}(x)=-\vert x^{\downarrow\rhd}\cup x \vert >- \vert y^{\downarrow\rhd}\cup y \vert=C_{\alpha_i}(y).$ 
If 2) is true, then the definition of $U$ implies $U(y)\geq U(x^{\rhd}_i)=\alpha_i> U(x).$
The definition of $C_{\alpha_i}(x)$ and the positivity of $U$ yield $C_{\alpha_i}(x)=\vert x^{\downarrow\rhd}\cup x \vert >- \vert y^{\downarrow\rhd}\cup y \vert=C_{\alpha_i}(y).$ 
Finally, if 3) is verified, then then the definition of $U$ implies $U(x^{\rhd}_i)=\alpha_i>U(x)>U(y).$
Using again the definition of $C_{\alpha_i}(x)$ we conclude that 
$C_{\alpha_i}(x)=\vert x^{\downarrow\rhd}\cup x \vert > \vert y^{\downarrow\rhd}\cup y \vert=C_{\alpha_i}(y).$

Assume now that $C_{\alpha_i}(x)>C_{\alpha_i}(y)$ for some $i\in\{0,\cdots,\vert X\vert-1\}$ and $x,y\in X$.
The definition of $C_{\alpha_i}$ and the fact that $U$ is always positive imply that three mutually exclusive cases are possible: (1 $U(y)>U(x)\geq \alpha_i$, (2 $U(y)\geq \alpha_i> U(x)$, or (3 $\alpha_i>U(x)>U(y)$.
Assume that (1 holds.
The definitions of $U$ and $\alpha_i$ yield either $y\rhd x\equiv x^{\rhd}_i$ or $y\rhd x\rhd x^{\rhd}_i$.
By Definition~\ref{DEF:compromise} we conclude that $x\rhd_i y$.
If (2 is true, then the definitions of $U$ and $\alpha_i$ yield either $y \equiv x^{\rhd}_i \rhd x$ or $y\rhd x^{\rhd}_i \rhd x$.
Definition~\ref{DEF:compromise} implies that that $x\rhd_i y$.
Finally, if (3 is verified, then the definitions of $U$ and $\alpha_i$ imply that $x^{\rhd}_i\rhd x\rhd y$.
 Definition~\ref{DEF:compromise} yields $x\rhd_i y$.

Since $(\rhd,Pr)$ is a justification by compromise of $\rho$, for any $i\in\{0,\cdots,\vert X\vert-1\}$ such that $Pr(\rhd_i)>0$ we have that  $C_{\alpha_i}$ represents $\rhd_i$, and $Pr(\rhd_i)=P(\alpha_i)$, we conclude that 

$$\rho(x,A)=\sum_{\alpha_i\in\mathcal{T}\colon x=\max_{y\in A}C_{\alpha_i}(y)}P(\alpha_i).$$

\noindent(ii)$(\Longrightarrow)$(i).
Let $\rhd\in\mathsf{LO}(X)$ be defined by $x\rhd y$ if $U(x)>U(y)$, for any $x,y\in X$.
Since $U$ is injective, it is straightforward to show that $\rhd$ is a linear order.
Moreover, for any $\alpha\in\mathcal{T}$ such that $P(\alpha)>0$, let $i$ be $\vert\{ x\in X\colon U(x)\geq \alpha\}\vert$ if $\vert\{ x\in X\colon U(x)\geq \alpha\}\vert\leq \vert X\vert-1$ or $\vert\{ x\in X\colon U(x)\geq \alpha\}\vert-1$ if $\vert\{ x\in X\colon U(x)\geq \alpha\}\vert=\vert X\vert$.
Note that $i\in\{0,\cdots,\vert X\vert\}.$ 
Thus, define $\rhd_i$ by $x\,\rhd_i\,y$ if $C_{\alpha}(x)>C_{\alpha}(y)$ for any $x,y\in X.$\footnote{Note that $\rhd_i$ is well defined for any $i\in\{0,\cdots,\vert X\vert-1\}$, since $C_{\alpha^{\prime}}(x)>C_{\alpha^{\prime}}(y)$  if and only if $C_{\alpha^{\prime\prime}}(x)>C_{\alpha^{\prime\prime}}(y)$ for any $\alpha^{\prime},\alpha^{\prime\prime}$ such that $\vert\{ x\in X\,\vert\, U(x)\geq \alpha^{\prime}\}\vert=\vert X\vert-1$, $\vert\{ x\in X\,\vert\,U(x)\geq \alpha^{\prime\prime}\}\vert=\vert X\vert$, and $x,y\in X$.}
Since $U$ is injective, each $C_{\alpha}$ is injective, and thus, it is straightforward to show that $\rhd_i$ is a linear order.
We now show that $\rhd_i$ is the $i$-th compromise on $\rhd$.
To see this, given $x,y\in X$, without loss of generality two mutually exclusive cases are possible: either $x,y\in X\setminus X^{\rhd}_i$ or $x\in X^{\rhd}_i$ and $y\in X$.
If the former is true, then the definitions of $\rhd$ and $i$ imply that $\alpha>U(x)>U(y).$
The definition of $C_{\alpha}$ implies that $C_{\alpha}(x)=U(x)>U(y)=C_{\alpha}(y).$
The definition of $\rhd_i$ yields $x\rhd_i y$.
If the latter case holds, the definitions of $\rhd$ and $i$ yield either $U(x)>U(y)\geq \alpha$ or $U(x)\geq \alpha > U(y).$   
The properties of $U$ and the definition of $C_{\alpha}$ imply that respectively either $C_{\alpha}(y)=-U(y)>-U(x)=C_{\alpha}(x)$ or $C_{\alpha}(y)=U(y)>-U(x)=C_{\alpha}(x).$ 
The definition of $\rhd_i$ yields $y\rhd_i x$.
Thus, by Definition~\ref{DEF:compromise}, $\rhd_i$ is the $i$-th compromise on $\rhd$.
Now, let $Pr\in\Delta(\mathsf{Comp}(\rhd))$ be the probability distribution defined by $Pr(\rhd_i)=\sum_{\alpha\in \mathcal{T}^{\,\prime}}P(\alpha)$ where $\mathcal{T}^{\,\prime}\subseteq \mathcal{T}$ is the set $\{\alpha\in\mathcal{T}\,\vert\,x\,\rhd_i y\,\text{if and only if}\, C_{\alpha}(x)>C_{\alpha}(y)\}$.\footnote{Since $P\in\Delta(\mathcal{T})$ is a probability distribution,  $Pr\in\Delta(\mathsf{Comp}(\rhd))$ is a probability distribution.}
Condition (ii) of the statement of Lemma~\ref{LEM:utility_representation_compromise_based_RUMs}, the fact that, for any $i\in\{0,\cdots,\vert X\vert-1\}$, $Pr(\rhd_i)=\sum_{\alpha\in \mathcal{T}^{\,\prime}}P(\alpha)$ with $\mathcal{T}^{\,\prime}=\{\alpha\in\mathcal{T}\,\vert\,C_{\alpha}\,\text{represent}\,\rhd_i\}$ and $\rhd_i$ is a compromise on $\rhd$, imply that

$$\rho(x,A)=\sum_{\rhd_{i}\in \mathsf{Comp}(\rhd)\colon x=\max(A,\rhd_i)}Pr(\rhd_i)$$
	holds for any $A\in\X$ and $x\in A.$
Finally, note also that it has already been proved that conditions (1), (2), and (3) of the statement of Lemma~\ref{LEM:utility_representation_compromise_based_RUMs} hold.}

\medskip
\noindent \textbf{\large Proof of Lemma~\ref{LEM:equality_definitions}}.
We need some preliminary results.
\begin{lemma}~\label{LEM:compromise-based_distortions_indices}
Given a finite set $X$, 	consider distinct indices $h,j\in\{1,\cdots\vert X\vert\}.$
	The following are equivalent:
	\begin{itemize}
		\item $x^{\rhd}_{h}\in x^{\uparrow\rhd}_{j}$,
		\item For any $k	\in\{0,\cdots,\vert X\vert-1\}$, $k<h$ if and only if $x_{h}^{\rhd}\rhd_k x^{\rhd}_j.$ 
	\end{itemize}  
\end{lemma}
\begin{proof}
This result is an immediate consequence of Definition~\ref{DEF:compromise}.	
\end{proof}

Lemma~\ref{LEM:compromise-based_distortions_indices} yields the following corollary.

\begin{corollary}\label{COR:compromise-based_distortions_indices_upper_lower}
	Let $X$ be a finite set, and consider indices $h,j\in\{1,\cdots,\vert X\vert\}$, and $k\in\{0,\cdots,\vert X\vert-1\}$ such that $h\neq j$.
	If $x^{\rhd}_h\in x_j^{\uparrow\rhd}$, then $x^{\rhd}_h \rhd_k x^{\rhd}_j$ if $k\leq h-1,$ and $ x^{\rhd}_j  \rhd_k x^{\rhd}_h$ if $k>h-1$.
	If $x^{\rhd}_l\in x_j^{\downarrow\rhd}$, then $x^{\rhd}_j\rhd_k x^{\rhd}_l$ if $k\leq j-1$, and $ x^{\rhd}_l \rhd_k x^{\rhd}_j$ if $k> j-1.$
\end{corollary}

We are now ready to prove Lemma~\ref{LEM:equality_definitions}. Consider a linear order $\rhd\in \mathsf{LO}(X)$, a $Pr$ over $\mathsf{Comp}(\rhd)$, a menu $A\in \X$, and an item $x\in A$ such that $x=x^{\rhd}_j$ for some  $1\leq j\leq \vert X\vert $.
Four cases are possible:
\begin{enumerate}
\item[(1)] $A_{x_j^{\uparrow\rhd}}\neq \es$ and $A_{x_j^{\downarrow\rhd}}=\es$,
	\item[(2)] $A_{x_j^{\uparrow\rhd}}\neq \es$ and $A_{x_j^{\downarrow\rhd}}\neq\es$,
	\item[(3)] $A_{x_j^{\uparrow\rhd}}= \es$ and $A_{x_j^{\downarrow\rhd}}=\es$, 
	\item[(4)] $A_{x_j^{\uparrow\rhd}}= \es$ and $A_{x_j^{\downarrow\rhd}}\neq\es$.
\end{enumerate}
If case (1) holds, by Definition~\ref{DEF:Stochastic_lack_of_confidence} and Corollary~\ref{COR:compromise-based_distortions_indices_upper_lower} I have that
\begin{align*}
\sum_{\rhd_{i}\in \mathsf{Comp}\colon x=\max(A,\rhd_i)}Pr(\rhd_i)=& \sum_{k\leq j-1} Pr(\rhd_{k})-\sum_{k< g\colon x^{\rhd}_g=\min\left(A_{x_j^{\uparrow\rhd}},\,\rhd\right)} Pr(\rhd_k)+\sum_{k\geq j} Pr(\rhd_{k}).	
\end{align*}
If case (2) holds, by Definition~\ref{DEF:Stochastic_lack_of_confidence} and Corollary~\ref{COR:compromise-based_distortions_indices_upper_lower} Ihave that
\begin{align*}
\sum_{\rhd_{i}\in \mathsf{Comp}(\rhd)\colon x=\max(A,\rhd_i)}Pr(\rhd_i) =& \sum_{k\leq j-1} Pr(\rhd_{k})-\sum_{k< g\colon x^{\rhd}_g=\min\left(A_{x_j^{\uparrow\rhd}},\,\rhd\right)} Pr(\rhd_k).	
\end{align*}
If case (3) holds, by Definition~\ref{DEF:Stochastic_lack_of_confidence} and Corollary~\ref{COR:compromise-based_distortions_indices_upper_lower} I have that
\begin{align*}
\sum_{\rhd_{i}\in \mathsf{Comp}(\rhd)\colon x=\max(A,\rhd_i)}Pr(\rhd_i)=& \sum_{k\leq j-1} Pr(\rhd_{k})+\sum_{k\geq j} Pr(\rhd_{k})=1.	
\end{align*}
Finally, if case (4) holds, by Definition~\ref{DEF:Stochastic_lack_of_confidence} and Corollary~\ref{COR:compromise-based_distortions_indices_upper_lower} I have that
\begin{align*}
\sum_{\rhd_{i}\in \mathsf{Comp}(\rhd)\colon x=\max(A,\rhd_i)}Pr(\rhd_i)= \sum_{k\leq j-1} Pr(\rhd_{k}).	
\end{align*}
Thus, the equality 
\begin{align*} 
\sum_{\rhd_{i}\in \mathsf{Comp}(\rhd)\colon x=\max(A,\rhd_i)}Pr(\rhd_i)=&\sum_{k\leq j-1} Pr(\rhd_{k})-	 \mathbf{1}_{\left\{A_{x_j^{\uparrow\rhd}}\neq \es\right\}}\sum_{k< g\colon x^{\rhd}_g=\min\left(A_{x_j^{\uparrow\rhd}},\,\rhd\right)} Pr(\rhd_k)+
\\&\mathbf{1}_{\left\{A_{x_j^{\downarrow\rhd}}=\es\right\}}\sum_{k\geq j} Pr(\rhd_{k})
\end{align*}
holds for each of the four cases above.
\qed
\smallskip

\noindent \textbf{\large Proof of Theorem~\ref{THM:compromise-based_stochastic_choices_characterization}}.
(\textit{\textbf{Only if part}}). Assume that $\rho\colon X\times \X\to [0,1]$ is a compromise-based RUM, and there is a pair $(\rhd,Pr)$ that justifies by compromise $\rho$.
Corollary~\ref{COR:nec_cond_compromise-based_choices_general} and Corollary~\ref{COR:identification_distortions_probability} imply that $\rhd$ composes $\rho$.

(\textit{\textbf{If part}}). Assume that some linear order $\rhd\in~\mathsf{LO}(X)$ composes $\rho\colon X\times \X\to [0,1]$. 
Let $Pr$ be the probability distribution over $\textsf{Harm}(\rhd)$ such that $Pr(\rhd_i)=\rho(x^{\rhd}_{i+1},X)$ for any $\rhd_i\in\textsf{Harm}(\rhd)$.
Note that, since $\sum_{j=1}^{\vert X\vert} \rho(x_j^{\rhd},X)=1$, we have that $\sum_{i=0}^{\vert X\vert-1} Pr(\rhd_i)=1$.
Moreover, since $\rhd$ composes $\rho$, I have that 
$$
\rho(x^{\rhd}_j,A)=\sum_{k\leq j} \rho(x^{\rhd}_{k},X)-\mathbf{1}_{\left\{A_{x_j^{\uparrow\rhd}}\neq \es\right\}}\sum_{k\leq g\colon x^{\rhd}_g=\min\left(A_{x_j^{\uparrow\rhd}},\,\rhd\right)}\rho(x^{\rhd}_k,X)+
\mathbf{1}_{\left\{A_{x_j^{\downarrow\rhd}}=\es\right\}}\sum_{k> j} \rho(x^{\rhd}_{k},X) 
$$
for any menu $A$, and any $1\leq j\leq \vert X\vert$ such that $x^{\rhd}_j\in A$.
Since $Pr(\rhd_i)=\rho(x^{\rhd}_{i+1},X)$ for any $0\leq i\leq \vert X\vert-1$, or, equivalently, $Pr(\rhd_{j-1})=\rho(x^{\rhd}_{j},X)$ for any $1\leq j\leq \vert X\vert $,  I obtain that  

$$	\rho(x^{\rhd}_j,A)= \sum_{k\leq j-1} Pr(\rhd_{k})-\mathbf{1}_{\left\{A_{x_j^{\uparrow\rhd}}\neq \es\right\}}\sum_{k< g\colon x^{\rhd}_g=\min\left(A_{x_j^{\uparrow\rhd}},\,\rhd\right)}Pr(\rhd_k)+
	\mathbf{1}_{\left\{A_{x_j^{\downarrow\rhd}}=\es\right\}}\sum_{k\geq j} Pr(\rhd_{k})$$
for any $A\in\X$, and any $1\leq j\leq \vert X\vert$ such that $x^{\rhd}_j\in A$.
Corollary~\ref{COR:nec_cond_compromise-based_choices_general} yields that $(\rhd,Pr)$ justifies by compromise $\rho.$
\qed
\smallskip

{\noindent\textbf{\large Proof of Theorem~\ref{THM:characterization_compromise_based_RUMs_exogenous_preference}}. (i)$(\Longleftrightarrow)$(ii). Proved in the proof of Theorem~\ref{THM:compromise-based_stochastic_choices_characterization}. 

\noindent (ii)$(\Longrightarrow)$(iii)
To show that persistent compromise holds, assume that there is $A\in\X$, and $x,y\in A$ such that $x\rhd y$ and there is no $z\in A$ satisfying $x\rhd z\rhd y$.
Assume that $x=x^{\rhd}_s$ and $y=x^{\rhd}_{t}$, where $s,t\in\{1,\cdots,\vert X\vert\}$ and $s<t$.
Since $\rhd$ composes $\rho$, Definition~\ref{DEF:ordered_composition}, $y\in A$, and $x\rhd y$ yield 

$$\rho(x^{\rhd}_s,A)= \sum_{k\leq s} \rho(x^{\rhd}_{k},X)-\mathbf{1}_{\left\{A_{x_s^{\uparrow\rhd}}\neq \es\right\}}\sum_{k\leq g\colon x^{\rhd}_g=\min\left(A_{x_s^{\uparrow\rhd}},\,\rhd\right)}\rho(x^{\rhd}_k,X).$$ 

Moreover, Definition~\ref{DEF:ordered_composition}, $x\in A$, $x\rhd y$, and the absence of any $z\in A$ such that $x\rhd z\rhd y$ imply
$$\rho(x^{\rhd}_t,A)= \sum_{s<k\leq t} \rho(x^{\rhd}_{k},X)+ 
	\mathbf{1}_{\left\{A_{x_t^{\downarrow\rhd}}=\es\right\}}\sum_{k> t} \rho(x^{\rhd}_{k},X).$$

Note  that

\begin{align}\label{EQ:equality_s_t}
 \sum_{k\leq t} \rho(x^{\rhd}_{k},X)= \sum_{k\leq s} \rho(x^{\rhd}_{k},X)+ \sum_{s<k\leq t} \rho(x^{\rhd}_{k},X)
\end{align}

holds.
Moreover, the fact that $x^{\rhd}_s=\min\left(A_{x_t^{\uparrow\rhd}},\,\rhd\right)$ yields  

\begin{align}\label{EQ:equality_A_t_A_s_t}
\sum_{k\leq g\colon x^{\rhd}_g=\min\left(A_{x_s^{\uparrow\rhd}},\,\rhd\right)}\rho(x^{\rhd}_k,X)=\sum_{k\leq g\colon x^{\rhd}_g=\min\left({B}_{x_t^{\uparrow\rhd}},\,\rhd\right)}\rho(x^{\rhd}_k,X)
\end{align}
 
where $B=A\setminus x^{\rhd}_s$. 
Finally, since $x\rhd y$, we also have that  

\begin{align}\label{EQ:equality_A_t_A_s_t_second}
A_{x_t^{\downarrow\rhd}}=B_{x_t^{\downarrow\rhd}}.
\end{align}

Definition~\ref{DEF:ordered_composition}, and equalities~\ref{EQ:equality_s_t}, \ref{EQ:equality_A_t_A_s_t}, and \ref{EQ:equality_A_t_A_s_t_second} imply that

\begin{align*}
\rho(x^{\rhd}_t,A\setminus x^{\rhd}_s)=&\sum_{k\leq s} \rho(x^{\rhd}_{k},X)+\sum_{s<k\leq t} \rho(x^{\rhd}_{k},X)-\mathbf{1}_{\left\{A_{x_s^{\uparrow\rhd}}\neq \es\right\}}\sum_{k\leq g\colon x^{\rhd}_g=\min\left(A_{x_s^{\uparrow\rhd}},\,\rhd\right)}\rho(x^{\rhd}_k,X)+\\ 
	&\mathbf{1}_{\left\{A_{x_t^{\downarrow\rhd}}=\es\right\}}\sum_{k> t} \rho(x^{\rhd}_{k},X)=\rho(x^{\rhd}_s,A)+\rho(x^{\rhd}_t,A).
	\end{align*}
We now prove that minimal rewards is verified.
Suppose that there are $A\in\X$, and $x,y\in A$ such that $x=\min(A,\rhd)$ and there is no $z\in A$ satisfying $y\rhd z\rhd x$.
Assume that $y=x^{\rhd}_s$ and $x=x^{\rhd}_{t}$, where $s,t\in\{1,\cdots,\vert X\vert\}$ and $s<t$.
Definition~\ref{DEF:ordered_composition}, $x\in A$, and $x=\min(A,\rhd)$ we have that 

$$\rho(x^{\rhd}_s,A)= \sum_{k\leq s} \rho(x^{\rhd}_{k},X)-\mathbf{1}_{\left\{A_{x_s^{\uparrow\rhd}}\neq \es\right\}}\sum_{k\leq g\colon x^{\rhd}_g=\min\left(A_{x_s^{\uparrow\rhd}},\,\rhd\right)}\rho(x^{\rhd}_k,X).$$

Definition~\ref{DEF:ordered_composition}, $y\in A$, the absence of any $z\in A$ such that $y\rhd z\rhd x$, and the fact that $x=\min(A,\rhd)$ yield
 
$$\rho(x^{\rhd}_t,A)= \sum_{s<k\leq t} \rho(x^{\rhd}_{k},X)+\sum_{k> t} \rho(x^{\rhd}_{k},X).$$

Note that, since there is no $z\in A$ such that $y\rhd z\rhd x$, we must have that 

\begin{align}\label{EQ:equality_third}
B_{x_s^{\downarrow\rhd}}=\es
\end{align}
  
where $B=A\setminus x^{\rhd}_t$.
Note also that, since $x\rhd y$, the equality

\begin{equation}\label{EQ:equality_third_new}
	A_{x_t^{\uparrow\rhd}}=B_{x_t^{\uparrow\rhd}}.
\end{equation}

Moreover, 

\begin{align}\label{EQ:equality_fourth}
\sum_{k>s} \rho(x^{\rhd}_{k},X)=\sum_{s<k\leq t} \rho(x^{\rhd}_{k},X)+\sum_{k> t} \rho(x^{\rhd}_{k},X)
\end{align}

holds.
Definition~\ref{DEF:ordered_composition}, and equalities \ref{EQ:equality_third},  \ref{EQ:equality_third_new}, \ref{EQ:equality_fourth} imply that 

\begin{align*}
\rho(x^{\rhd}_s,A\setminus x^{\rhd}_t)= & \sum_{k\leq s} \rho(x^{\rhd}_{k},X)-\mathbf{1}_{\left\{A_{x_s^{\uparrow\rhd}}\neq \es\right\}}\sum_{k\leq g\colon x^{\rhd}_g=\min\left(A_{x_s^{\uparrow\rhd}},\,\rhd\right)}\rho(x^{\rhd}_k,X)+\\
&\sum_{s<k\leq t} \rho(x^{\rhd}_{k},X)+\sum_{k> t} \rho(x^{\rhd}_{k},X)=\rho(x^{\rhd}_s,A)+\rho(x^{\rhd}_t,A).
\end{align*}
Finally, to see that compromise-based centrality holds, assume that there is $A\in\X$ and $x,y\in A$ such that 
\begin{enumerate}
		\item[1)] there is $z\in A$ such that $y\rhd x\rhd z$, or 
\item[2)] there is $z\in A$ such that $y\rhd z\rhd x$, or
\item[3)] there is $z\in A$ such that $x\rhd z\rhd y$. 
\end{enumerate}

If conditions 1) or 2) are true, then assume that $y=x^{\rhd}_s$ and $x=x^{\rhd}_{t}$, where $s,t\in\{1,\cdots,\vert X\vert\}$ and $s<t$.
Definition~\ref{DEF:ordered_composition}, $y\rhd x$ and $x\in A$ imply that 

$$\rho(x^{\rhd}_s,A)= \sum_{k\leq s} \rho(x^{\rhd}_{k},X)-\mathbf{1}_{\left\{A_{x_s^{\uparrow\rhd}}\neq \es\right\}}\sum_{k\leq g\colon x^{\rhd}_g=\min\left(A_{x_s^{\uparrow\rhd}},\,\rhd\right)}\rho(x^{\rhd}_k,X).$$

Since $y\rhd x$, we have that

\begin{align}\label{EQ:equality_fifth}
 A_{x_s^{\uparrow\rhd}}=B_{x_s^{\uparrow\rhd}}
\end{align}

with $B=A\setminus x^{\rhd}_t.$
Moreover, $z\in A$ implies that 

\begin{align}\label{EQ:equality_sixth}
B_{x_s^{\downarrow\rhd}}\neq \es.
\end{align}

Definition~\ref{DEF:ordered_composition}, and the equalities~\ref{EQ:equality_fifth} and \ref{EQ:equality_sixth} yield

 $$\rho(x^{\rhd}_s,A\setminus x^{\rhd}_t)= \sum_{k\leq s} \rho(x^{\rhd}_{k},X)-\mathbf{1}_{\left\{A_{x_s^{\uparrow\rhd}}\neq \es\right\}}\sum_{k\leq g\colon x^{\rhd}_g=\min\left(A_{x_s^{\uparrow\rhd}},\,\rhd\right)}\rho(x^{\rhd}_k,X)=\rho(x^{\rhd}_s,A).$$

If (3) is true, then suppose that $x=x^{\rhd}_s$ and $y=x^{\rhd}_{t}$, where $s,t\in\{1,\cdots,\vert X\vert\}$ and $s<t$.
Definition~\ref{DEF:ordered_composition}, $x\rhd y$, and $x\in A$ imply that

$$\rho(x^{\rhd}_t,A)=\sum_{k\leq t} \rho(x^{\rhd}_{k},X)-\mathbf{1}_{\left\{A_{x_t^{\uparrow\rhd}}\neq \es\right\}}\sum_{k\leq g\colon x^{\rhd}_g=\min\left(A_{x_t^{\uparrow\rhd}},\,\rhd\right)}\rho(x^{\rhd}_k,X)+ \mathbf{1}_{\left\{A_{x_t^{\downarrow\rhd}}=\es\right\}}\sum_{k> t} \rho(x^{\rhd}_{k},X).$$

Since $z\in A$, and $x\rhd z\rhd y$ we must have that

\begin{align}\label{EQ:equality_sevent_new}
A_{x_t^{\uparrow\rhd}}\neq\es \Longleftrightarrow B_{x_t^{\uparrow\rhd}}\neq\es 
\end{align}

and

\begin{align}\label{EQ:equality_seventh}
\min\left(A_{x_t^{\uparrow\rhd}},\rhd\right)=	\min\left(B_{x_t^{\uparrow\rhd}},\rhd\right),
\end{align}

with $B=A\setminus x^{\rhd}_s$.
Finally, note that since $x\rhd y$, the equality 

\begin{align}\label{EQ:equality_eight}
A_{x_t^{\downarrow\rhd}}=B_{x_t^{\downarrow\rhd}}
\end{align}

Definition~\ref{DEF:ordered_composition}, implication \ref{EQ:equality_sevent_new}, and equalities~\ref{EQ:equality_seventh} and \ref{EQ:equality_eight} yield

\begin{align*}
\rho(x^{\rhd}_t,A\setminus x^{\rhd}_s)=&\sum_{k\leq t} \rho(x^{\rhd}_{k},X)-\mathbf{1}_{\left\{A_{x_t^{\uparrow\rhd}}\neq \es\right\}}\sum_{k\leq g\colon x^{\rhd}_g=\min\left(A_{x_t^{\uparrow\rhd}},\,\rhd\right)}\rho(x^{\rhd}_k,X)+\\
&\mathbf{1}_{\left\{A_{x_t^{\downarrow\rhd}}=\es\right\}}\sum_{k> t} \rho(x^{\rhd}_{k},X)=
\rho(x^{\rhd}_t,A).
\end{align*}

\noindent (iii)$(\Longrightarrow)$(i)
Let $Pr\in\Delta(\mathsf{Comp}(\rhd))$ be the probability distribution defined by $Pr(\rhd_i)=\rho(x^{\rhd}_{i+1},X)$, for any $i\in\{0,\cdots,\vert X\vert-1\}.$
Note that, since $\sum_{j=1}^{\vert X\vert} \rho(x_j^{\rhd},X)=1$, we have that $\sum_{i=0}^{\vert X\vert-1} Pr(\rhd_i)=1$.
Consider a menu $A\in\X$, and some $y\in A$ such $y\equiv x^{\rhd}_j$ for some $j\in\{1,\cdots,\vert X\vert\}.$
Assume also that $\vert X\vert\geq 3$.\footnote{When $\vert X\vert=2$ the proof is straightforward.}
Three cases, mutually exclusive, are possible:
\begin{itemize}
	\item[(1] $y=\max(X,\rhd)$, or 
	\item[(2] $y\neq\max(X,\rhd) $ and $y\neq \min(X,\rhd)$, or
	\item[(3] $y=\min(X,\rhd).$
\end{itemize}

If $(1$ is true, then $j=1$ and $x^{\rhd}_j\equiv x^{\rhd}_1$.
If $\vert A\vert=1$, then
 $$\rho(x^{\rhd}_1,A)=1=\sum_{\rhd_{i}\in \mathsf{Comp}(\rhd)\colon y=\max(A,\rhd_i)}Pr(\rhd_i).$$

Assume now that $\vert A\vert\geq 2.$
Note that, since $y=\max(X,\rhd)$, for any $B^{\prime}\in\X$ such that $B^{\prime}\supset A$ there are always some $x,z\in B$ such that either $y\rhd x\rhd z$ or $y\rhd z\rhd x$ is verified. 
Thus, since cases (1) or (2) of the hypothesis of minimal rewards, presented in Definition~\ref{DEF:characterizing_behavioral_properties}, holds for any $B^{\prime}\in\X$ such that $B^{\prime}\supset A$, we can apply this property recursively to conclude that $\rho(x^{\rhd}_1,A)=\rho(x^{\rhd}_1,X)=Pr(\rhd_i)$.
By Definition \ref{DEF:compromise} we have that $Pr(\rhd_i)=\sum_{\rhd_{i}\in \mathsf{Comp}(\rhd)\colon x^{\rhd}_1=\max(A,\rhd_i)}Pr(\rhd_i).$
Thus, we conclude that 

$$\rho(y,A)=\rho(x^{\rhd}_1,X)=\sum_{\rhd_{i}\in \mathsf{Comp}(\rhd)\colon x^{\rhd}_1=\max(A,\rhd_i)}Pr(\rhd_i)=\sum_{\rhd_{i}\in \mathsf{Comp}(\rhd)\colon y=\max(A,\rhd_i)}Pr(\rhd_i).$$

Assume now that $(2$ holds.
Then consider the following mutually exclusive subcases:

\begin{itemize}
	\item[(2(a)] $A_{\,x_j^{\uparrow\,\rhd}}\neq \es$ and $A_{\,x_j^{\downarrow\,\rhd}}= \es$, or  
		\item[(2(b)]  $A_{\,x_j^{\uparrow\,\rhd}}\neq \es$ and $A_{\,x_j^{\downarrow\,\rhd}}\neq \es$, or 
	\item[(2(c)] $A_{\,x_j^{\uparrow\,\rhd}}= \es$ and $A_{\,x_j^{\downarrow\,\rhd}}\neq \es$, or
	\item[(2(d)]$A_{\,x_j^{\uparrow\,\rhd}}= \es$ and $A_{\,x_j^{\downarrow\,\rhd}}=\es$.
\end{itemize}

If case (2(a is verified, then consider $x^{\rhd}_g=\min{\left(A_{x_j^{\uparrow\rhd}},\,\rhd\right)}$, which, since $A_{\,x_j^{\uparrow\,\rhd}}\neq \es$, exists.
Note that, for any $B\in\X$ such that $B\supset A$  and there is $x^{\rhd}_f\in B$ such that $x^{\rhd}_f\rhd x^{\rhd}_g\rhd x^{\rhd}_j$, condition (3) of compromise-based centrality, stated in Definition~\ref{DEF:characterizing_behavioral_properties} is satisfied.
Thus, we can apply recursively such property to conclude that $\rho(x^{\rhd}_j,A^{\prime})=\rho(x^{\rhd}_j,X),$ 
where the set $A^{\prime}\supseteq A$ is $X\setminus\{x^{\rhd}_f\in X,\,\vert,\, x^{\rhd}_f \rhd x^{\rhd}_g\wedge x^{\rhd}_f\not\in A\}$.

Consider now $x^{\rhd}_h=\min(\{x^{\rhd}_l\in A^{\prime}\,\vert\, x^{\rhd}_g\rhd x^{\rhd}_l\rhd x^{\rhd}_j\},\rhd),$ if any.
Since persistent compromise applies we get $\rho(x^{\rhd}_j, A^{\prime}\setminus x^{\rhd}_h)=\rho(x^{\rhd}_j,A^{\prime})+\rho(x^{\rhd}_h,A^{\prime})$.
The same property, applied iteratively, yields $\rho(x^{\rhd}_j,A^{\prime\prime})=\sum_{g<k\leq j}\rho(x^{\rhd}_k,A^{\prime})$, with $A^{\prime\prime}=A^{\prime}\setminus \{x^{\rhd}_l\in A^{\prime}\,\vert\, x^{\rhd}_g\rhd x^{\rhd}_l\rhd x^{\rhd}_j\}$ and  $A^{\prime}\supseteq A^{\prime\prime}\supseteq A$.
Moreover, for any $D\supset A^{\prime}$, and $x^{\rhd}_k\in A^{\prime}$ with $g<k\leq j$, there is $z\in D$ such that $z\rhd x^{\rhd}_g\rhd x^{\rhd}_k$.
We can recursively apply condition (3) of compromise-based centrality to conclude that $\rho(x_{k}^{\rhd},A^{\prime})=\rho(x^{\rhd}_k,X)$, for any $g< k\leq j.$ 
Thus,  $\rho(x^{\rhd}_j,A^{\prime\prime})=\sum_{g<k\leq j}\rho(x^{\rhd}_k,X).$

Iteratively apply now condition (1) of compromise-based centrality to obtain that $\rho(x^{\rhd}_j,A^{\prime\prime\prime})=\rho(x^{\rhd}_j,A^{\prime\prime})=\sum_{g<k\leq j}\rho(x^{\rhd}_k,X)$, with $A^{\prime\prime\prime}=A^{\prime\prime}\setminus\{x^{\rhd}_l\,\vert\, j<l<\vert X\vert\}$, and $A^{\prime\prime}\supseteq A^{\prime\prime\prime} \supset A$.
Finally since $x^{\rhd}_j\in A^{\prime\prime\prime}$, minimal rewards yields $\rho(x^{\rhd}_j,A)=\rho\left(x^{\rhd}_j,A^{\prime\prime\prime}\setminus x^{\rhd}_{\vert X\vert}\right)=\rho(x^{\rhd}_j,A^{\prime\prime\prime})+\rho\left(x^{\rhd}_{\vert X\vert},A^{\prime\prime\prime}\right)=\sum_{g<k\leq j}\rho(x^{\rhd}_k,X)+\rho\left(x^{\rhd}_{\vert X\vert},A^{\prime\prime\prime}\right)$.
Since $x^{\rhd}_j\in A^{\prime\prime}$, for any $D\in\X$ such that $D\supset A^{\prime\prime}$ we can apply recursively condition (3) of compromise-based centrality to conclude that $\rho\left(x^{\rhd}_{l},A^{\prime\prime}\right)=\rho\left(x^{\rhd}_{l},X\right)$ for any $l\in\{j+1,\cdots,\vert X\vert\}.$
The same condition yields 

\begin{equation}\label{EQ:probabilities_pinning_down}
\rho\left(x^{\rhd}_{l},E\right)=\rho\left(x^{\rhd}_{l},A^{\prime\prime}\right)=\rho\left(x^{\rhd}_{l},X\right) 
\end{equation}

for any $l\in\{j+1,\cdots,\vert X\vert\}$ and $E\subseteq A^{\prime\prime}$ containing $x^{\rhd}_j$ and some $x^{\rhd}_p$ such that $x^{\rhd}_j\rhd x^{\rhd}_p\rhd x^{\rhd}_{l}.$    
Persistent compromise implies that 

$$\rho\left(x^{\rhd}_{\vert X\vert},A^{\prime\prime}\setminus{x^{\rhd}_{\vert X\vert-1}}\right)=\rho\left(x^{\rhd}_{\vert X\vert},A^{\prime\prime}\right)+\rho\left(x^{\rhd}_{\vert X\vert-1},A^{\prime\prime}\right)=\rho\left(x^{\rhd}_{\vert X\vert},X\right)+\rho\left(x^{\rhd}_{\vert X\vert-1},X\right).$$

Apply again persistent compromise to conclude that

 \begin{align*}
 	& \rho\left(x^{\rhd}_{\vert X\vert},\left(A^{\prime\prime}\setminus{x^{\rhd}_{\vert X\vert-1}}\right)\setminus x^{\rhd}_{\vert X\vert-2}\right)= \rho\left(x^{\rhd}_{\vert X\vert},A^{\prime\prime}\setminus{x^{\rhd}_{\vert X\vert-1}}\right)+\rho\left(x^{\rhd}_{\vert X\vert-2},A^{\prime\prime}\setminus{x^{\rhd}_{\vert X\vert-1}}\right)=\\ &\rho\left(x^{\rhd}_{\vert X\vert},X\right)+\rho\left(x^{\rhd}_{\vert X\vert-1},X\right)+\rho\left(x^{\rhd}_{\vert X\vert-2},A^{\prime\prime}\setminus{x^{\rhd}_{\vert X\vert-1}}\right) .
 	\end{align*}

Since $x^{\rhd}_{\vert X\vert}\in A^{\prime\prime}\setminus x^{\rhd}_{\vert X\vert-1}$, we apply condition (1) of compromise-based centrality to get $\rho\left(x^{\rhd}_{\vert X\vert-2},A^{\prime\prime}\setminus{x^{\rhd}_{\vert X\vert-1}}\right)=\rho\left(x^{\rhd}_{\vert X\vert-2},A^{\prime\prime}\right)=\rho\left(x^{\rhd}_{\vert X\vert-2},X\right).$ 
Thus, $\rho\left(x^{\rhd}_{\vert X\vert},\left(A^{\prime\prime}\setminus{x^{\rhd}_{\vert X\vert-1}}\right)\setminus x^{\rhd}_{\vert X\vert-2}\right)=\rho\left(x^{\rhd}_{\vert X\vert},X\right)+\rho\left(x^{\rhd}_{\vert X\vert-1},X\right)+\rho\left(x^{\rhd}_{\vert X\vert-2},X\right).$
Since equality~\ref{EQ:probabilities_pinning_down} holds, this argument can be iterated until we obtain that $\rho\left(x^{\rhd}_{\vert X\vert},A^{\prime\prime\prime}\right)=\sum_{j< k\leq \vert X\vert}\rho(x^{\rhd}_k,X).$
Thus, we conclude that 

$$\rho(x^{\rhd}_j,A)=\rho(x^{\rhd}_j,A^{\prime\prime\prime})+\rho\left(x^{\rhd}_{\vert X\vert},A^{\prime\prime\prime}\right)=\sum_{g< k\leq j}\rho(x^{\rhd}_k,X)+\sum_{j< k\leq \vert X\vert}\rho(x^{\rhd}_k,X)=\sum_{g< k}\rho(x^{\rhd}_k,X).$$

The definition of $Pr$ implies that  $\sum_{g< k}\rho(x^{\rhd}_k,X)=\sum_{g\leq i\leq \vert X\vert-1}Pr(\rhd_{i})$.
Finally, Definition~\ref{DEF:compromise} yields $\sum_{g\leq i\leq \vert X\vert-1}Pr(\rhd_{i})=\sum_{\rhd_i\in\mathsf{Comp}(\rhd)\colon x^{\rhd}_j=\max(A,\rhd_i)}Pr(\rhd_{i})$.
We conclude that 

$$\rho(y,A)=\rho(x^{\rhd}_j,A)=\sum_{\rhd_i\in\mathsf{Comp}(\rhd)\colon x^{\rhd}_j=\max(A,\rhd_i)}Pr(\rhd_{i})=\sum_{\rhd_i\in\mathsf{Comp}(\rhd)\colon y=\max(A,\rhd_i)}Pr(\rhd_{i}).$$

If case (2(b) holds, since $A_{\,x_j^{\uparrow\,\rhd}}\neq \es$, we can use the same argument of case 2)(a) to conclude that $\rho(x^{\rhd}_j,A^{\prime\prime})=\sum_{g<k\leq j}\rho(x^{\rhd}_k,X),$ where $A^{\prime\prime}=A^{\prime}\setminus \{x^{\rhd}_l\in A^{\prime}\,\vert\, x^{\rhd}_g\rhd x^{\rhd}_l\rhd x^{\rhd}_j\}$ and $A^{\prime}=X\setminus\{x^{\rhd}_f\in X\,\vert\, x^{\rhd}_f \rhd x^{\rhd}_g\wedge x^{\rhd}_f\not\in A\}$.
Since $A_{\,x_j^{\downarrow\,\rhd}}\neq \es$ and $A\subseteq A^{\prime\prime}$, we can apply iteratively condition (1) or condition  (2) of compromise-based centrality, or both to conclude that $\rho(x^{\rhd}_j,A)=\rho(x^{\rhd}_j,A^{\prime\prime})=\sum_{g<k\leq j}\rho(x^{\rhd}_k,X).$
The definition of $Pr$ implies that  $\sum_{g< k\leq j}\rho(x^{\rhd}_k,X)=\sum_{g\leq i<j}Pr(\rhd_{i})$.
Finally, Definition \ref{DEF:compromise} yields $\sum_{g\leq i<j}Pr(\rhd_{i})=\sum_{\rhd_i\in\mathsf{Comp}(\rhd)\colon x^{\rhd}_j=\max(A,\rhd_i)}Pr(\rhd_{i})$.
We can conclude that 

$$\rho(y,A)=\rho(x^{\rhd}_j,A)=\sum_{\rhd_i\in\mathsf{Comp}(\rhd)\colon x^{\rhd}_j=\max(A,\rhd_i)}Pr(\rhd_{i})=\sum_{\rhd_i\in\mathsf{Comp}(\rhd)\colon y=\max(A,\rhd_i)}Pr(\rhd_{i}).$$

Assume case (2(c) is verified.
Condition (1) of compromise-based centrality implies that for any $A\subseteq X$ and each $x^{\rhd}_d$ such that $x^{\rhd}_d, x^{\rhd}_j\in A,$ and there is $x^{\rhd}_f\in A$ satisfying $x^{\rhd}_d \rhd x^{\rhd}_f \rhd x^{\rhd}_j$, we have

\begin{equation}\label{EQ:equality_compromise_based_centrality_other_items_ground_set}
	\rho(x^{\rhd}_d,A)=\rho(x^{\rhd}_d,A\setminus x^{\rhd}_f).
\end{equation}
 
Observe now that persistent compromise implies that  $\rho(x^{\rhd}_j,X\setminus x^{\rhd}_{j-1})=\rho(x^{\rhd}_j,X)+\rho(x^{\rhd}_{j-1},X).$
Apply again persistent compromise to get 

\begin{align*}
&\rho(x^{\rhd}_j,(X\setminus x^{\rhd}_{j-1})\setminus x^{\rhd}_{j-2})=\rho(x^{\rhd}_j,X\setminus x^{\rhd}_{j-1})+\rho(x^{\rhd}_{j-2},X\setminus x^{\rhd}_{j-1})=\\ &\rho(x^{\rhd}_j,X)+\rho(x^{\rhd}_{j-1},X)+\rho(x^{\rhd}_{j-2},X\setminus x^{\rhd}_{j-1}).
\end{align*}

Apply equality~\ref{EQ:equality_compromise_based_centrality_other_items_ground_set} to get $\rho(x^{\rhd}_{j-2},X\setminus x^{\rhd}_{j-1})=\rho(x^{\rhd}_{j-2},X).$
We can use iteratively this argument to conclude that $\rho(x^{\rhd}_j,A^{\prime})=\sum_{k\leq j}\rho(x^{\rhd}_k,X)$, with $A^{\prime}=X\setminus x^{\uparrow\,\rhd}_j$ and $A^{\prime}\supseteq A.$
Since $A_{\,x_j^{\downarrow\,\rhd}}\neq \es$ and $A\subseteq A^{\prime}$, we can apply iteratively condition (1), or condition (2) of compromise-based centrality, or both, to conclude that $\rho(x^{\rhd}_j,A)=\rho(x^{\rhd}_j,A^{\prime})=\sum_{k\leq j}\rho(x^{\rhd}_k,X).$ 
The definition of $Pr$ implies that  $\sum_{k\leq j}\rho(x^{\rhd}_k,X)=\sum_{i<j}Pr(\rhd_{i})$.
Finally, Definition \ref{DEF:compromise} yields $\sum_{i<j}Pr(\rhd_{i})=\sum_{\rhd_i\in\mathsf{Comp}(\rhd)\colon x^{\rhd}_j=\max(A,\rhd_i)}Pr(\rhd_{i})$.
We can conclude that 

$$\rho(y,A)=\rho(x^{\rhd}_j,A)=\sum_{\rhd_i\in\mathsf{Comp}(\rhd)\colon x^{\rhd}_j=\max(A,\rhd_i)}Pr(\rhd_{i})=\sum_{\rhd_i\in\mathsf{Comp}(\rhd)\colon y=\max(A,\rhd_i)}Pr(\rhd_{i}).$$

If case (2(d) is true, then $A=\{x^{\rhd}_j\}.$
Definition~\ref{DEF:stochastic_choice} implies that $\rho(x^{\rhd}_j,A)=1.$
Definition~\ref{DEF:compromise} implies that $\rho(y,A)=\sum_{\rhd_i\in\mathsf{Comp}(\rhd)\colon y=\max(A,\rhd_i)}Pr(\rhd_{i}).$

Finally, if case 3) is verified, then $j=\vert X\vert$ and $x^{\rhd}_j\equiv x^{\rhd}_{\vert X\vert}.$
If $\vert A\vert=1$, then
 $$\rho(x^{\rhd}_1,A)=1=\sum_{\rhd_{i}\in \mathsf{Comp}(\rhd)\colon y=\max(A,\rhd_i)}Pr(\rhd_i).$$

Assume now that $\vert A\vert\geq 2.$ 
Thus, $A_{\,x_{\vert X\vert}^{\uparrow\,\rhd}}\neq \es$. 
We can use the same argument adopted in cases 2)(a) and 2)(b) to conclude that $\rho(x^{\rhd}_j,A^{\prime\prime})=\sum_{g<k\leq \vert X\vert}\rho(x^{\rhd}_k,X),$ where $A^{\prime\prime}=A^{\prime}\setminus \{x^{\rhd}_l\in A^{\prime}\,\vert\, x^{\rhd}_g\rhd x^{\rhd}_l\rhd x^{\rhd}_{\vert X\vert}\}$ and $A^{\prime}=X\setminus\{x^{\rhd}_f\in X\,\vert\, x^{\rhd}_f \rhd x^{\rhd}_g\wedge x^{\rhd}_f\not\in A\}$.
Note that $A^{\prime\prime}=A.$
We conclude that $\rho\left(x^{\rhd}_{\vert X\vert},A\right)=\sum_{g<k\leq \vert X\vert}\rho(x^{\rhd}_k,X).$
The definition of $Pr$ implies that  $\sum_{k\leq \vert X\vert}\rho(x^{\rhd}_k,X)=\sum_{i\leq \vert X\vert-1}Pr(\rhd_{i})$.
Finally, Definition \ref{DEF:compromise} yields $\sum_{i\leq\vert X\vert-1}Pr(\rhd_{i})=\sum_{\rhd_i\in\mathsf{Comp}(\rhd)\colon x^{\rhd}_j=\max(A,\rhd_i)}Pr(\rhd_{i})$.
We can conclude that 

$$\rho(y,A)=\rho(x^{\rhd}_{\vert X\vert},A)=\sum_{\rhd_i\in\mathsf{Comp}(\rhd)\colon x^{\rhd}_{\vert X\vert}=\max(A,\rhd_i)}Pr(\rhd_{i})=\sum_{\rhd_i\in\mathsf{Comp}(\rhd)\colon y=\max(A,\rhd_i)}Pr(\rhd_{i}).$$

\qed}

\smallskip

\noindent \textbf{\large Proof of Lemma~\ref{LEM:uniqueness_up_to_probability_distribution_preliminary}}.
Some preliminary notation.
Given a linear order $\rhd\in\mathsf{LO}(X)$, and an item $x\in X$, I denote by $x^{\uparrow\rhd}$ the set $\{y\in X\setminus\{x\}\,\vert\, y\rhd x\}$.
We use the following result.
\begin{theorem}[\citealt{Turansick2022}]\label{THM:Turansick_identification_theorem}
Assume that $\rho\colon X\times\X\to[0,1]$ is a RUM, and $Pr\in\Delta(\mathsf{LO}(X))$ justifies $\rho$.
Then $Pr$ is the unique probability distribution that explains $\rho$ if and only if there is no pair of linear orders $\rhd$, $\rhd^{\prime}$ that satisfy the following conditions.
\begin{enumerate}[\rm(i)]
	\item $Pr(\rhd)>0$ and $Pr(\rhd^{\prime})>0;$ 
	\item there are $x,y,z\in X$ such that
	\begin{enumerate}
	\item $x,y\rhd z$, and $x,y\rhd^{\prime} z$,
	\item $x\neq y$,
	\item $\left(z^{\uparrow\rhd}\cup z\right)\neq \left(z^{\uparrow\rhd^{\prime}}\cup z\right)$, 
	\item $\left(x^{\uparrow\rhd}\cup x\right)=\left(y^{\uparrow\rhd^{\prime}}\cup y\right)$.
	\end{enumerate}
\end{enumerate}
\end{theorem}  
We call conditions (i) and (ii) of Theorem~\ref{THM:Turansick_identification_theorem} the \textit{Turansick's conditions}. Assume now toward a contradiction that $p\colon X\times\X\to[0,1]$ is a compromise-based RUM, and that there are two distinct probability distributions $Pr, Pr^{\prime}\in\Delta(\mathsf{LO}(X))$ that rationalize $\rho$.
Thus, the Turansick's conditions hold.
Since $\rho$ is a compromise-based RUM, I can conclude that there is $\rhd\in\mathsf{LO}(X)$, and distinct $i,j\in\{0,\cdots,\vert X\vert -1\}$ such that
\begin{enumerate}[\rm(i)]
\item
	 
 $Pr(\rhd_i)>0,$ and $Pr(\rhd_j)>0$;
 \item there are $x,y,z\in X$ such that
 \begin{enumerate}
 \item $x,y\,\rhd_i z$, $x,y\rhd_j z$,
 \item $x\neq y$,
 \item $\left(z^{\uparrow\rhd_i}\cup z\right)\neq \left(z^{\uparrow\rhd_j}\cup z\right)$,
 \item  $\left(x^{\uparrow\rhd_i}\cup x\right)=\left(y^{\uparrow\rhd_j}\cup y\right)$.
 \end{enumerate}
 \end{enumerate}
Moreover, without loss of generality, assume that $i<j$, and $y\rhd x$, and that the items $y,x,z$  occupy respectively the $k$-th, $l$-th, and $p$-th position in $X$, with respect to $\rhd$, that is $y=x^{\rhd}_k, x=x^{\rhd}_l$, and $z=x^{\rhd}_p$, with $1\leq k<l\leq\vert X\vert $.
Definition~\ref{DEF:compromise} and condition (ii)(c) yields $i< p$.
By Definition~\ref{DEF:compromise} I know also that, looking at the position of $y$ and $x$ with respect to $\rhd_i$ and $\rhd_j$, three mutually exclusive cases are possible: (1) $j<k$, and,  as a consequence, $y\rhd_i x,$ $y\,\rhd_j x$, (2)  $k\leq i$, and, as a consequence,  $x\rhd_i y$, $x\rhd_j y$, or (3) $i< k\leq j$, and, as a consequence,  $y\rhd_i x$, $x\rhd_j y$.
If case (1) holds, I obtain that $x\in (x^{\uparrow\rhd_i}\cup x)$, but $x\not\in \left(y^{\uparrow\rhd_j}\cup y\right)$, which contradicts condition (ii)(d).
If case (2) holds, I obtain that $y\not\in(x^{\uparrow\rhd_i}\cup x),$ and  $y\in  \left(y^{\uparrow\rhd_j}\cup y\right)$, which again contradicts condition (ii)(d). 
Finally, if case (3) holds, three subcases are possible: (3)(a) $z\rhd y\rhd x$, or, equivalently, $p<k<l$, (3)(b) $y\rhd x\rhd z$, or, equivalently, $k<l<p$, or (3)(c) $y\rhd z\rhd x,$ or, equivalently $k<p<l$.
 If subcase (3)(a) holds, since $i<p$, I obtain that $z\rhd_i y\rhd_i x$, which contradicts condition (ii)(a). 
If subcase (3)(b) holds, Definition~\ref{DEF:compromise} implies that either  (3)(b)$^{\prime}$ $l\leq j$, and thus $y\rhd_i x\rhd_i z$, and $z\rhd_j x\rhd_j y$, or (3)(b)$^{\prime\prime}$ $j<l$, and thus $y\rhd_i x\rhd_i z$, and $ x\rhd_j  z\rhd_j y$.
However, (3)(b)$^{\prime}$ and (3)(b)$^{\prime\prime}$  contradict  condition (ii)(a).
Finally, if subcase (3)(c) holds then by Definition~\ref{DEF:compromise} I have that $y \rhd_i z\rhd_i x$, which contradicts condition (ii)(a).
We conclude that the Turansick's conditions do not hold, and that the probability distribution $Pr$ is the unique one that rationalizes $\rho$. 
\qed
\smallskip

\noindent \textbf{\large Proof of Theorem~\ref{THM:stochastic_self_punishment_identification}}.
(i)$(\Longrightarrow)$(ii).
We prove this by contrapositive, that is, I show that, given a stochastic choice $\rho\colon X\times\X\to[0,1]$ on a set of cardinality $\vert X\vert\geq 3,$ if at least one of the conditions
\begin{itemize}
	\item[(1)] $\rhd$ composes $\rho$,
	\item[(2)] $\vert X^*\vert\geq 3,$ or $\vert X^*\vert=2$, 
 and $\min(X,\rhd)\not\in X^*,$	
\end{itemize}
fails, then $(\rhd,Pr_{\rho,\rhd})$ is not the unique justification by compromise of $\rho.$
If (1) does not hold, then by Corollary~\ref{COR:derivation_probability_from_dataset} $(\rhd,Pr_{\rho,\rhd})$ is not a justification by compromise of $\rho.$

If (2) does not hold (and (1) holds),  then either $\vert X^*\vert=1$, or $\vert X^*\vert=2,$ and $\min(X,\rhd)\in X^*.$

If $\vert X^*\vert=1$, then since (1) holds, $\rhd$ composes $\rho,$ and by Corollary~\ref{COR:derivation_probability_from_dataset} $(\rhd,Pr_{\rho,\rhd})$ is a justification by compromise of $\rho.$
Moreover, since  $\vert X^*\vert=1,$ then there is $j\in\{1,\cdots,\vert X\vert\}$ such that $x^{\rhd}_{j}\in X^{*}.$
Corollary~\ref{COR:identification_distortions_probability} implies that $Pr_{\rho,\rhd}(\rhd_{j-1})=1,$ and $Pr_{\rho,\rhd}(\rhd_{k})=0,$ for every $k\in \{0,\vert X\vert-1\}$ distinct from $j-1$.
By Corollary~\ref{LEM:uniqueness_up_to_probability_distribution}, it is enough to show that there is $\rhd^{\prime}\not\equiv \rhd$ such that $\rhd_{j-1}\in\mathsf{Comp}(\rhd^{\prime}).$
Thus, let $\rhd^{\prime}\in\mathsf{LO}(X)$ be defined by $x^{\rhd^{\prime}}_h=x^{\rhd}_{h},$ for all $1\leq h <j$, and  $x^{\rhd^{\prime}}_h=x^{\rhd}_{\vert X\vert+j-h}$ for any $\vert X\vert \geq h\geq j.$ 
We claim that $\rhd_{j-1}\equiv \rhd^{\prime}_{\vert X\vert -1}.$
To see this, note that for any $x,y\in X$ such that, without loss of generality,  $x\rhd_{j-1} y$ holds, by  Definition~\ref{DEF:compromise} two cases are possible:
\begin{itemize}
	\item[1)] $y=x^{\rhd}_{k},$ and $x=x^{\rhd}_l$, $1\leq k<l,$ and $k<j$ or 
	\item[2)] $x=x^{\rhd}_{k},$ and $y=x^{\rhd}_l$, and $1\leq j\leq k<l\leq \vert X\vert$.
\end{itemize} 

If 1) holds, then the definition of $\rhd^{\prime}$ implies $y\rhd^{\prime} x$. 
Apply again Definition~\ref{DEF:compromise}  to obtain $x\rhd^{\prime}_{\vert X\vert-1} y.$   
If 2) holds, the definition of $\rhd^{\prime}$ implies $y\rhd^{\prime} x$.
Apply again Definition~\ref{DEF:compromise}  to obtain $x\rhd^{\prime}_{\vert X\vert-1} y.$

If $\vert X^*\vert=2$ and $\min(X,\rhd)\in X^*,$ since (1) holds, $\rhd$ composes $\rho,$ and by Corollary~\ref{COR:derivation_probability_from_dataset} $(\rhd,Pr_{\rho,\rhd})$ is a justification by compromise of $\rho.$
Moreover, since $\vert X^*\vert=2,$ and $\min(X,\rhd)\in X^{*},$ then there is $j\in\{1,\cdots,\vert X\vert-1\}$ such that $x^{\rhd}_{j}\in X^{*}.$
Corollary~\ref{COR:identification_distortions_probability} implies that $Pr_{\rho,\rhd}(\rhd_{j-1})>0,$ and $Pr_{\rho,\rhd}(\rhd_{\vert X\vert-1})>0.$
By Corollary~\ref{LEM:uniqueness_up_to_probability_distribution}, it is enough to show that there is $\rhd^{\prime}\not\equiv \rhd$ such that $\{\rhd_{j-1},\rhd_{\vert X\vert-1}\}\subset\mathsf{Comp}(\rhd^{\prime}).$
Thus, let $\rhd^{\prime}\in\mathsf{LO}(X)$ be defined, as before, by $x^{\rhd^{\prime}}_h=x^{\rhd}_{h},$ for all $1\leq h <j$, and  $x^{\rhd^{\prime}}_h=x^{\rhd}_{\vert X\vert+j-h}$ for any $h\geq j.$ 
We claim that $\rhd_{j-1}\equiv \rhd^{\prime}_{\vert X\vert -1},$ and $\rhd_{\vert X\vert -1}\equiv \rhd^{\prime}_{j-1}.$
To show that $\rhd_{j-1}\equiv \rhd^{\prime}_{\vert X\vert -1},$ note that for any $x,y\in X$ such that, without loss of generality,  $x\rhd_{j-1} y$ holds, by  Definition~\ref{DEF:compromise} two cases are possible:
\begin{itemize}
	\item[1)] $y=x^{\rhd}_{k},$ and $x=x^{\rhd}_l$, $1\leq k<l,$ and $k<j$ or 
	\item[2)] $x=x^{\rhd}_{k},$ and $y=x^{\rhd}_l$, and $1\leq j\leq k<l\leq \vert X\vert$.
\end{itemize} 

If 1) holds, then the definition of $\rhd^{\prime}$ implies $y\rhd^{\prime} x$. 
Apply again Definition~\ref{DEF:compromise}  to obtain $x\rhd^{\prime}_{\vert X\vert-1} y.$   
If 2) holds, the definition of $\rhd^{\prime}$ implies $y\rhd^{\prime} x$.
Apply again Definition~\ref{DEF:compromise}  to obtain $x\rhd^{\prime}_{\vert X\vert-1} y.$ 

To show that  $\rhd_{\vert X\vert -1}\equiv \rhd^{\prime}_{j-1},$ note that for any $x,y\in X$ such that, without loss of generality,  $x\rhd_{\vert X\vert -1} y$ holds, by  Definition~\ref{DEF:compromise} I have $y\rhd x.$
Consider the following mutually exclusive subcases:

\begin{itemize}
	\item[3)] $y=x^{\rhd}_{k},$ and $x=x^{\rhd}_l$, $1\leq k<l,$ and $k<j$ or 
	\item[4)] $y=x^{\rhd}_{k},$ and $x=x^{\rhd}_l$, and $1\leq j\leq k<l\leq \vert X\vert$.
\end{itemize} 

If 3) holds, then the definition of $\rhd^{\prime}$ implies $y\rhd^{\prime} x$.
Apply Definition~\ref{DEF:compromise}  to conclude that $x\rhd^{\prime}_{i-1} y$ holds.
If 4) holds, the definition of $\rhd^{\prime}$ implies $x\rhd^{\prime} y$.
Apply Definition~\ref{DEF:compromise} to conclude that $x\rhd^{\prime}_{j-1} y.$

\medskip
\noindent(i)$(\Longleftarrow)$(ii).
We need some preliminary results.

\begin{lemma}\label{LEM:unique_true_preference_three_item}
Assume that $\vert X\vert \geq 3$, and there is $\rhd\in\mathsf{LO}(X)$ and $i,j,k\in\{0,\cdots,\vert X\vert-1\}$ such that $i< j< k,$ $x^{\rhd_i}_{1}=x^{\rhd_j}_{\vert X\vert}=x^{\rhd_k}_{\vert X\vert}.$
Then $\{\rhd_i,\rhd_j,\rhd_k\}\not\subseteq \mathsf{Comp}(\rhd^{\prime})$ for any $\rhd^{\prime}\not\equiv\rhd.$ 
\end{lemma}

\begin{proof}
Assume toward a contradiction that there are $\rhd,\rhd^{\prime}\in\mathsf{LO}(X)$ such that $x^{\rhd_i}_{1}=x^{\rhd_j}_{\vert X\vert}=x^{\rhd_k}_{\vert X\vert}$,  $\{\rhd_i,\rhd_j,\rhd_k\}\subseteq \mathsf{Comp}(\rhd)$, and $\{\rhd_i,\rhd_j,\rhd_k\}\subseteq \mathsf{Comp}(\rhd^{\prime})$.
Thus, there are $l,m,n\in\{0,\cdots,\vert X\vert-1\}$ such that $\rhd_i\equiv \rhd^{\prime}_l, \rhd_j\equiv \rhd^{\prime}_m,$ and $\rhd_k\equiv \rhd^{\prime}_n.$
Since $\{\rhd_i,\rhd_j,\rhd_k\}\subseteq \mathsf{Comp}(\rhd)$ and $x^{\rhd_i}_{1}=x^{\rhd_j}_{\vert X\vert}=x^{\rhd_k}_{\vert X\vert}$, by Definition~\ref{DEF:compromise} I have that $i=0.$
Since $\{\rhd^{\prime}_l,\rhd^{\prime}_m,\rhd^{\prime}_n\}\subseteq \mathsf{Comp}(\rhd^{\prime}),$ and $x^{\rhd_l}_{1}=x^{\rhd_m}_{\vert X\vert}=x^{\rhd_n}_{\vert X\vert}$ by Definition~\ref{DEF:compromise} I have that $l=0.$
Thus, I must have that $\rhd^{\prime}\equiv\rhd^{\prime}_0\equiv\rhd_0\equiv	\rhd$, which is false.
\end{proof}

\begin{lemma}\label{LEM:distortions_equals_same_index}
Assume that $\vert X\vert\geq 2,$ and there are $\rhd\in\mathsf{LO}(X)$ and $i,j\in\{0,\cdots,\vert X\vert-1\}$	such that $\rhd_i\equiv\rhd_j.$
Then $i=j.$
\end{lemma}
\begin{proof}
	We proof the result by contrapositive.
	Thus, assume without loss of generality that $i<j,$ for some $i,j\in\{0,\cdots\vert X\vert-1\}.$
	Let $\rhd\in\mathsf{LO}(X)$ be some linear order on $X.$
	By Definition~\ref{DEF:compromise} I have that $x^{\rhd_i}_{\vert X\vert-i}=x^{\rhd}_{\vert X\vert},$ and $x^{\rhd_{j}}_{\vert X\vert-i}=x^{\rhd}_{i+1}.$
	Since $i<j\leq \vert X\vert-1,$ I obtain that $x^{\rhd}_{\vert X\vert}\neq x^{\rhd}_{i+1}$ and that $x^{\rhd_i}_{\vert X\vert-i}\neq x^{\rhd_{j}}_{\vert X\vert-i}.$
	Thus, $\rhd_{i}\not\equiv \rhd_j.$
\end{proof}

\begin{lemma}\label{LEM:harful_distortions_same_indices_same_preference}
	Assume that $\vert X\vert\geq 2,$ and there are $\rhd,\rhd^{\prime}\in\mathsf{LO}(X)$ and $i\in\{0,\cdots,\vert X\vert-1\}$ such that $\rhd_i\equiv\rhd^{\prime}_i.$
	Then $\rhd\equiv\rhd^{\prime}.$
\end{lemma}

\begin{proof}
	We prove this result by contrapositive.
	Assume that $\rhd\not\equiv \rhd^{\prime}.$
		Thus, there is $y\in X$ s.t. $y=x^{\rhd}_{k},$ and $y=x^{\rhd^{\prime}}_{l},$ with $k,l\in\{1,\cdots,\vert X\vert\},$ and $k\neq l.$
	Consider some $i\in\{0,\cdots\vert X\vert-1\}.$
	By Definition~\ref{DEF:compromise} I have that $x^{\rhd_{i}}_{\vert X\vert-k+1}=y,$ but $x^{\rhd^{\prime}_{i}}_{\vert X\vert-k+1}\neq y,$ which implies that $\rhd_{i}\neq\rhd^{\prime}_{i}.$
\end{proof}

\begin{lemma}\label{LEM:two_compromise-based_distortions_two_preferences_inversion_indices}
Assume that $\vert X\vert\geq 2$, and there are $i,j,k,l\in\{0,\cdots,\vert X\vert-1\},$ and $\rhd,\rhd^{\prime}\in\mathsf{LO}(X)$ such that $0<i<j$, $\rhd\not\equiv \rhd^{\prime},$ $\rhd_{i}\equiv\rhd^{\prime}_{l},$ and $\rhd_{j}\equiv\rhd^{\prime}_k$.
Then $k<l,$ $i=k$, $l=j=\vert X\vert-1$.	
\end{lemma}

\begin{proof}
Note that $k\neq l,$ otherwise I would obtain $\rhd^{\prime}_{k}\equiv\rhd^{\prime}_{l},$ which by Lemma \ref{LEM:distortions_equals_same_index} implies that $\rhd_{i}\equiv \rhd_j$ and $i=j,$ which is false.
Thus two cases are possible:
\begin{enumerate}[\rm(i)]
	\item $l<k$, or 
	\item $l>k.$
\end{enumerate}
If case (i) holds, note that I must have that $l\neq i,$ otherwise I would get $\rhd_{i}\equiv \rhd^{\prime}_i,$ which implies  by Lemma~\ref{LEM:harful_distortions_same_indices_same_preference} that $\rhd\equiv \rhd^{\prime},$ which is false.
Thus, consider $\min\{i,l\}.$
There are two subcases:
\begin{itemize}
	\item[(i)(a)] $i=\min\{i,l\}$, or 
	\item[(i)(b)] $l=\min\{i,l\}.$
\end{itemize}
Assume subcase (i)(a) holds.
By Definition~\ref{DEF:compromise}	we have that $x^{\rhd_i}_{\vert X\vert-i}=x^{\rhd}_{\vert X\vert}.$ 
Definition~\ref{DEF:compromise} and $i<j$ imply that $x^{\rhd_j}_{\vert X\vert-i}=x^{\rhd}_{i+1}$.
Note also that, since $i<j\leq \vert X\vert-1$, I can conclude that $x^{\rhd}_{i+1}\neq x^{\rhd}_{\vert X\vert}.$
Definition~\ref{DEF:compromise}  and $i<l<k$ yield $x^{\rhd^{\prime}_l}_{\vert X\vert-i}=x^{\rhd^{\prime}}_{i+1}=x^{\rhd^{\prime}_{k}}_{\vert X\vert-i}.$
We obtain that $x^{\rhd_i}_{\vert X\vert-i}\neq x^{\rhd_j}_{\vert X\vert-i}$ and $x^{\rhd^{\prime}_{l}}_{\vert X\vert-i}=x^{\rhd^{\prime}_k}_{\vert X\vert-i}.$
However note that, since $\rhd_{i}=\rhd^{\prime}_l$ and $\rhd_{j}=\rhd^{\prime}_{k},$ I must have that $x^{\rhd_i}_{\vert X\vert-i}=x^{\rhd^{\prime}_{l}}_{\vert X\vert-i},$ and $x^{\rhd_j}_{\vert X\vert-i}=x^{\rhd^{\prime}_k}_{\vert X\vert-i},$ which imply that $x^{\rhd_{i}}_{\vert X\vert-i}\neq x^{\rhd_{i}}_{\vert X\vert-i},$ a contradiction.

Assume that subcase (i)(b) holds.
By Definition~\ref{DEF:compromise} I have that $x^{\rhd^{\prime}_l}_{\vert X\vert-l}=x^{\rhd^{\prime}}_{\vert X\vert}.$
Definition~\ref{DEF:compromise} and $l<k$ imply $x^{\rhd^{\prime}_k}_{\vert X\vert-l}=x^{\rhd}_{l+1}.$
Note also that, since $l<k\leq \vert X\vert-1$, I can conclude that $x^{\rhd}_{l+1}\neq x_{\vert X\vert}.$
Definition~\ref{DEF:compromise}  and $l<i<j$ yields $x^{\rhd_{i}}_{\vert X\vert-l}=x^{\rhd}_{l+1}=x^{\rhd_{j}}_{\vert X\vert-l}.$
	We obtain that $x^{\rhd^{\prime}_l}_{\vert X\vert-l}\neq x^{\rhd^{\prime}_k}_{\vert X\vert-l}$  and $x^{\rhd_{i}}_{\vert X\vert-l}=x^{\rhd_{j}}_{\vert X\vert-l}.$
	However note that, since $\rhd_{i}=\rhd^{\prime}_l$ and $\rhd_{j}=\rhd^{\prime}_{k}$, I must have that  $x^{\rhd^{\prime}_{l}}_{\vert X\vert-l}=x^{\rhd_{i}}_{\vert X\vert-l},$ and $x^{\rhd^{\prime}_{k}}_{\vert X\vert-l}=x^{\rhd_{j}}_{\vert X\vert-l},$ which imply that $x^{\rhd^{\prime}_{l}}_{\vert X\vert-l}\neq x^{\rhd^{\prime}_{l}}_{\vert X\vert-l},$ a contradiction.

Since subcases (i)(a) and (i)(b) lead to a contradiction, I conclude that case (i) is false, and that (ii) holds, i.e., $l>k.$
To show that $i=k,$ consider the other two cases
\begin{itemize}
	\item[(ii)(a)] $i<k,$ and
	\item[(ii)(b)] $i>k.$
\end{itemize} 
Suppose that case (ii)(a) is true.
Definition~\ref{DEF:compromise} implies that $x^{\rhd_{i}}_{\vert X\vert -i}=x^{\rhd}_{\vert X\vert}.$
Definition~\ref{DEF:compromise} and $i<j$ yield $x^{\rhd_{j}}_{\vert X\vert -i}=x^{\rhd}_{i+1}.$
Note that, since $i<j\leq\vert X\vert-1,$ I conclude that $x^{\rhd}_{\vert X\vert}\neq x^{\rhd}_{i+1}.$
Definition~\ref{DEF:compromise} and $i<k<l$ imply that $x^{\rhd^{\prime}_k}_{\vert X\vert-i}=x^{\rhd^{\prime}}_{i+1}=x^{\rhd^{\prime}_l}_{\vert X\vert-i}.$
We obtain that $x^{\rhd_{i}}_{\vert X\vert -i}\neq x^{\rhd_{j}}_{\vert X\vert -i},$ and $x^{\rhd^{\prime}_k}_{\vert X\vert-i}=x^{\rhd^{\prime}_l}_{\vert X\vert-i}.$
However, since $\rhd_{i}=\rhd^{\prime}_l$ and $\rhd_{j}=\rhd^{\prime}_{k}$, I must have that $x^{\rhd_{i}}_{\vert X\vert-l}=x^{\rhd^{\prime}_{l}}_{\vert X\vert-l}=$, and $x^{\rhd_{j}}_{\vert X\vert-l}=x^{\rhd^{\prime}_{k}}_{\vert X\vert-l}$,  which imply that $x^{\rhd_{i}}_{\vert X\vert -i}\neq x^{\rhd_{i}}_{\vert X\vert -i},$ a contradiction.

Thus, suppose that (ii)(b) is true.
 Definition~\ref{DEF:compromise} implies that $x^{\rhd^{\prime}_{k}}_{\vert X\vert -k}=x^{\rhd^{\prime}}_{\vert X\vert}.$
 Definition~\ref{DEF:compromise} and $k<l$ yield $x^{\rhd^{\prime}_{l}}_{\vert X\vert -k}=x^{\rhd^{\prime}}_{k+1}.$
 Since $k<l\leq \vert X\vert-1,$ I must have that $x^{\rhd^{\prime}}_{\vert X\vert}\neq x^{\rhd^{\prime}}_{k+1}.$
 Definition~\ref{DEF:compromise} and $k<i<j$ imply that $x^{\rhd_i}_{\vert X\vert-k}=x^{\rhd}_{k+1}=x^{\rhd_j}_{\vert X\vert-k}.$
 I obtain that $x^{\rhd^{\prime}_{k}}_{\vert X\vert -k}\neq x^{\rhd^{\prime}_{l}}_{\vert X\vert -k},$ and $x^{\rhd_i}_{\vert X\vert-k}= x^{\rhd_j}_{\vert X\vert-k}.$
 However, since $\rhd_{i}=\rhd^{\prime}_l$ and $\rhd_{j}=\rhd^{\prime}_{k}$, I must have that $x^{\rhd^{\prime}_k}_{\vert X\vert-k}=x^{\rhd_j}_{\vert X\vert-k},$ and $x^{\rhd^{\prime}_{l}}_{\vert X\vert -k}=x^{\rhd_i}_{\vert X\vert-k},$ which imply that $x^{\rhd^{\prime}_k}_{\vert X\vert-k}\neq x^{\rhd^{\prime}_k}_{\vert X\vert-k},$ a contradiction.

Since (ii)(a) and (ii)(b) are false, I conclude that $i=k.$

We now show that $j=l=\vert X\vert-1$.
Definition~\ref{DEF:compromise}, $\rhd_i\equiv\rhd^{\prime}_l$, and $i<l$ imply that $x^{\rhd}_{\vert X\vert}=x^{\rhd_i}_{\vert X\vert-i}=x^{\rhd^{\prime}_l}_{\vert X\vert-i}=x^{\rhd^{\prime}}_{i+1}=x^{\rhd^{\prime}_i}_1.$ 
Definition~\ref{DEF:compromise}, $i=k$, $\rhd_{j}\equiv\rhd^{\prime}_k ,$ and $i<j$ yield $x^{\rhd^{\prime}}_{\vert X\vert}=x^{\rhd^{\prime}_i}_{\vert X\vert-i}=x^{\rhd_{j}}_{\vert X\vert-i}=x^{\rhd}_{i+1}=x^{\rhd_i}_1.$
We conclude that $x^{\rhd}_{\vert X\vert}=x^{\rhd^{\prime}_i}_1,$ and $x^{\rhd^{\prime}}_{\vert X\vert}=x^{\rhd_i}_1.$

Definition~\ref{DEF:compromise}, $\rhd_j\equiv \rhd^{\prime}_k,$ $i=k,$ and $x^{\rhd}_{\vert X\vert}=x^{\rhd^{\prime}_i}_1$ imply that $x^{\rhd}_{j+1}=x^{\rhd_j}_1=x^{\rhd^{\prime}_k}_1=x^{\rhd^{\prime}_i}_1=x^{\rhd}_{\vert X\vert}.$
By Lemma~\ref{LEM:distortions_equals_same_index} I obtain $\vert X\vert=j+1,$ which yields $j=\vert X\vert -1$.
Similarly, $x^{\rhd^{\prime}}_{\vert X\vert}=x^{\rhd_i}_1$, $\rhd_i\equiv \rhd^{\prime}_l,$ and Definition~\ref{DEF:compromise} imply that $x^{\rhd^{\prime}}_{\vert X\vert}=x^{\rhd_i}_1=x^{\rhd^{\prime}_l}_1=x^{\rhd^{\prime}}_{l+1}.$
By Lemma~\ref{LEM:distortions_equals_same_index} I obtain that $l=\vert X\vert-1,$ and $l=j.$ 
\end{proof}

\begin{lemma}\label{LEM:two_compromise-based_distortions_with_true_preference}
Assume that $\vert X\vert \geq 2$, and there is $\rhd\in\mathsf{LO}(X)$ and $i,j\in\{0,\cdots,\vert X\vert-1\}$ such that $i< j,$ $x^{\rhd_i}_{\vert X\vert}\neq x^{\rhd_j}_{\vert X\vert}.$
We have that $i=0$.
Moreover, if  $\{\rhd_i,\rhd_j\}\subseteq \mathsf{Comp}(\rhd^{\prime})$ for some $\rhd^{\prime}\not\equiv\rhd,$ then $j=\vert X\vert-1$, $\rhd\equiv\rhd^{\prime}_{\vert X\vert-1}$, and $\rhd_{\vert X\vert-1}\equiv \rhd^{\prime}$. 
\end{lemma}

\begin{proof}
Since $x^{\rhd_i}_{\vert X\vert}\neq x^{\rhd_j}_{\vert X\vert}$, and $i<j$,  Definition~\ref{DEF:compromise} yields $i=0,$ and $x^{\rhd}_{1}= x^{\rhd_j}_{\vert X\vert}.$
Since $\{\rhd,\rhd_j\}\subseteq \mathsf{Comp}(\rhd^{\prime})$,  there are $k,l\in\{0,\vert X\vert-1\}$ such that $\rhd \equiv \rhd^{\prime}_k$, and $\rhd_j\equiv \rhd^{\prime}_l$.

Definition~\ref{DEF:compromise}, $x^{\rhd^{\prime}_k}_{\vert X\vert}\neq x^{\rhd^{\prime}_l}_{\vert X\vert}$, $\rhd\not\equiv \rhd^{\prime},$ and $\rhd\equiv \rhd^{\prime}_k$ imply that  that $l=0,$ $\rhd_j\equiv \rhd^{\prime},$ and $x^{{\rhd^{\prime}}}_{1}=x^{\rhd}_{\vert X\vert}$. 
Moreover,  $\rhd\equiv \rhd^{\prime}_k$, Definition~\ref{DEF:compromise}, $i=0$, and $\rhd_j\equiv \rhd^{\prime}$ yield $x^{{\rhd^{\prime}_k}}_{1}=x^{{\rhd}}_{1}=x^{{\rhd_j}}_{\vert X\vert}=x^{{\rhd^{\prime}}}_{\vert X\vert}.$
We apply Definition~\ref{DEF:compromise} to conclude that $k=\vert X\vert-1,$ and thus, $\rhd\equiv \rhd^{\prime}_{\vert X\vert-1}.$
We also have that $\rhd_j\equiv \rhd^{\prime}$ and $x^{{\rhd^{\prime}}}_{1}=x^{\rhd}_{\vert X\vert}$  yield $x^{{\rhd_j}}_{1}=x^{{\rhd^{\prime}}}_{1}=x^{{\rhd}}_{\vert X\vert}$.
We  apply again Definition~\ref{DEF:compromise} to conclude that $j=\vert X\vert-1,$ and thus, $\rhd_{\vert X\vert-1}\equiv \rhd^{\prime}.$ 
\end{proof}

Given Corollary~\ref{COR:derivation_probability_from_dataset}, I can  assume toward a contradiction that $\rhd$ composes $\rho$, $\vert X^*\vert \geq 3,$  and either \begin{itemize}
	
	\item[$\star$] $(\rhd,Pr_{\rho,\rhd})$ is not a justification by compromise of $\rho$, or
	\item[$\star\star$] $(\rhd,Pr_{\rho,\rhd})$  a justification by compromise of $\rho$  and there is $\rhd^{\prime}\not\equiv \rhd$, such that $(\rhd^{\prime},Pr_{\rho,\rhd^{\prime}})$ is also a justification by compromise of $\rho$.
  
\end{itemize}

Condition $\star$ contradicts Corollary~\ref{COR:derivation_probability_from_dataset}.
Assume that $\star\star$ holds.
Since $(\rhd,Pr_{\rho,\rhd})$, and $\vert X^*\vert\geq 3$, by Corollary~\ref{COR:identification_distortions_probability} I know that there are $i,j,k\in\{0,\cdots,\vert X\vert-1\}$ such that $i<j<k$, $Pr_{\rho,\rhd}(\rhd_i)\neq 0$, $Pr_{\rho,\rhd}(\rhd_j)\neq 0$, $Pr_{\rho,\rhd}(\rhd_k)\neq 0$.
By Corollary~\ref{LEM:uniqueness_up_to_probability_distribution} $\{\rhd_i,\rhd_j,\rhd_k\}\subseteq\mathsf{Comp}(\rhd^{\prime})$.
Two cases are possible: 1) $0\in\{i,j,k\}$ or 2) $0\not\in\{i,j,k\}$.
If 1) holds, I have that $i=0$.
Definition~\ref{DEF:compromise} yields that $x^{\rhd_i}_{1}=x^{\rhd_j}_{\vert X\vert}=x^{\rhd_k}_{\vert X\vert}$.
Lemma~\ref{LEM:unique_true_preference_three_item} yields $\{\rhd_i,\rhd_j,\rhd_k\}\not\subseteq \mathsf{Comp}(\rhd^{\prime})$ for any $\rhd^{\prime}\not\equiv\rhd,$ a contradiction. 

If 2) holds, note that since $\{\rhd_{i},\rhd_{j},\rhd_{k}\}\subseteq \mathsf{Comp}(\rhd^{\prime}),$ there are $l,m,n\in\{0,\cdots,\vert X\vert-1\}$ such that for any $g\in\{i,j,k\}$ there is one and only one $h\in\{l,m,n\}$ for which $\rhd_{g}=\rhd^{\prime}_{h}.$
Note also that $\rhd_{l}\not\equiv \rhd_{m}\not\equiv \rhd_{n},$ and $\rhd_{l}\not\equiv\rhd_{n}.$
Consider the compromise $\rhd_{i}.$
By Lemma~\ref{LEM:two_compromise-based_distortions_two_preferences_inversion_indices} I  must have that $m=i=n,$ which yields $\rhd^{\prime}_{m}\equiv\rhd^{\prime}_n,$ which is false. 
Since conditions $\star$ and $\star\star$ lead to a contradiction, I conclude that when $\vert X^*\vert\geq 3$, and $\rhd$ composes $\rho$, the pair $(\rhd,Pr_{\rho,\rhd})$ is the unique justification by compromise of $\rho.$

Given Corollary~\ref{COR:derivation_probability_from_dataset}, I can assume toward a contradiction now that $\rhd$ composes $\rho$, $\vert X^*\vert= 2,$ $\min(X,\rhd)\not\in X^{*}$  and either \begin{itemize}
	
	\item[$\diamondsuit$] $(\rhd,Pr_{\rho,\rhd})$ is not a justification by compromise of $\rho$, or
	\item[$\diamondsuit\diamondsuit$] $(\rhd,Pr_{\rho,\rhd})$  a justification by compromise of $\rho$  and that there is $\rhd^{\prime}\not\equiv \rhd$ such that $(\rhd^{\prime},Pr_{\rho,\rhd^{\prime}})$ is also a justification by compromise of $\rho$.
  
\end{itemize}

Condition $\diamondsuit$ contradicts Corollary ~\ref{COR:derivation_probability_from_dataset}.
Thus assume that  $\diamondsuit\diamondsuit$ holds.
By Corollary~\ref{COR:identification_distortions_probability} I know that there are $i,j\in\{0,\cdots,\vert X\vert-2\}$ such that $i<j$, $Pr_{\rho,\rhd}(\rhd_i)> 0$, $Pr_{\rho,\rhd}(\rhd_j)> 0$, and $Pr_{\rho,\rhd}(\rhd_k)= 0$ for any $k\in\{0,\cdots,\vert X\vert-1\}\setminus\{i,j\}$.
By Corollary~\ref{LEM:uniqueness_up_to_probability_distribution} $\{\rhd_i,\rhd_j\}\subseteq \mathsf{Comp}(\rhd^{\prime})$. 
Two cases are possible: (1 $i=0$ or (2 $i>0.$

If (1 holds,
Definition~\ref{DEF:compromise} implies that $x^{\rhd_i}_{1}=x^{\rhd}_{1}=x^{\rhd_j}_{\vert X\vert}.$
Since $\rhd^{\prime}\not\equiv\rhd$, I apply Lemma~\ref{LEM:two_compromise-based_distortions_with_true_preference} to conclude that $\rhd\equiv \rhd^{\prime}_{\vert X\vert-1}$, and $\rhd_j\equiv\rhd_{\vert X\vert-1}\equiv\rhd^{\prime}.$
Lemma~\ref{LEM:distortions_equals_same_index} yields $j=\vert X\vert-1,$ a contradiction.
If (2 holds, then, since $\{\rhd_i,\rhd_j\}\subset\mathsf{Comp}(\rhd^{\prime})$, I can apply Lemma~\ref{LEM:two_compromise-based_distortions_two_preferences_inversion_indices} to conclude that $j=\vert X\vert-1,$ a contradiction. 
\qed
 
\noindent \textbf{\large Proof of Lemma~\ref{LEMMA:unique_two_justifications_two_items}}.
When I proved that condition (i) of Theorem~\ref{THM:stochastic_self_punishment_identification} implies condition (ii) of the same result I showed that if $\rhd\in\mathsf{LO}(X)$ composes $\rho\colon X\times\X\to[0,1] ,$ $\vert  X^{*} \vert=2$, $\min(X,\rhd)\in X^{*}$, and $\rho(x^{\rhd}_{j},X)> 0,$ for some $j\in\{0,\cdots,\vert X\vert-1\}$, then $(\rhd, Pr_{\rho,\rhd})$ and $(\rhd^{*j}, Pr_{\rho,\rhd^{*j}})$ are two justifications by compromise of $\rho$.
Moreover, I have that $\rhd_{j-1}\equiv \rhd^{*j}_{\vert X\vert -1},$ and $\rhd_{\vert X\vert -1}\equiv \rhd^{*}_{j-1}.$
By Lemma~\ref{LEM:uniqueness_up_to_probability_distribution_preliminary} I conclude that $Pr_{\rho,\rhd}(\rhd_{j-1})=Pr_{\rho,\rhd^{*j}}\left(\rhd^{*j}_{\vert X\vert-1}\right)> 0,$ and $Pr_{\rho,\rhd}(\rhd_{\vert X\vert-1})=Pr_{\rho,\rhd^{*j}}\left(\rhd^{*j}_{j-1}\right)> 0.$ 

Thus,  I are only left to show that $(\rhd,Pr_{\rho,\rhd})$, and $(\rhd^{*i},Pr_{\rho,\rhd^{*i}})$ are the only two distinct justifications by compromise of $\rho.$
By Corollary~\ref{COR:derivation_probability_from_dataset} it is enough to show that there is no $\rhd^{\prime}$ distinct from $\rhd$ and $\rhd^{*j}$ such that $(\rhd^{\prime},Pr_{\rho,\rhd^{\prime}})$ is a justification by compromise of $\rho.$
By Corollary~\ref{LEM:uniqueness_up_to_probability_distribution} I only have to prove that there is no $\rhd^{\prime}$ distinct from $\rhd$ and $\rhd^{*j}$ such that $\{\rhd_{j-1},\rhd_{\vert X\vert-1}\}=\{\rhd^{*}_{j-1}, \rhd^{*j}_{\vert X\vert -1}\}\subseteq \mathsf{Harm(\rhd^{\prime})}.$
To see this, assume toward a contradiction that there is $\rhd^{\prime}$ distinct from  $\rhd$ and $\rhd^{*j}$ such that  $\{\rhd_{j-1},\rhd_{\vert X\vert-1}\}=\{\rhd^{*j}_{j-1}, \rhd^{*j}_{\vert X\vert -1}\}\subseteq \mathsf{Harm(\rhd^{\prime})}.$
Two cases are possible:
\begin{itemize}
\item[(1)] $j=1$,
\item[(2)] $j\in\{2,\cdots,\vert X\vert-1\}.$
\end{itemize}
If (1) holds, then I have that $\{\rhd,\rhd_{\vert X\vert-1}\}\subseteq \mathsf{Comp}(\rhd^{*j}),$ and $\{\rhd,\rhd_{\vert X\vert-1}\}\subseteq \mathsf{Comp}(\rhd^{\prime}).$
We apply Lemma~\ref{LEM:two_compromise-based_distortions_with_true_preference} to conclude that $\rhd_{\vert X\vert-1}\equiv \rhd^{*j},$ and  $\rhd_{\vert X\vert-1}\equiv \rhd^{\prime},$ which yields $\rhd^{*j}\equiv \rhd^{\prime},$ a contradiction.

If (2) holds, then, since $\{\rhd_{j-1},\rhd_{\vert X\vert-1}\}\subseteq \mathsf{Comp}(\rhd^{\prime})$, I can apply Lemma~\ref{LEM:two_compromise-based_distortions_two_preferences_inversion_indices} to conclude that $\rhd_{j-1}\equiv \rhd^{\prime}_{\vert X\vert-1},$ and $\rhd_{\vert X\vert-1}\equiv\rhd^{\prime}_{j-1}.$
Since I already know that $\rhd_{j-1}\equiv \rhd^{*j}_{\vert X\vert -1},$ and $\rhd_{\vert X\vert-1}\equiv \rhd^{*j}_{j -1},$ I apply Lemma~\ref{LEM:harful_distortions_same_indices_same_preference} conclude that $\rhd^{\prime}\equiv \rhd^{*j},$ a contradiction.
\qed
\smallskip

\noindent \textbf{\large Proof of Lemma~\ref{LEMMA:unique_two_justifications_one_item}}.
Since $\rhd$ composes $\rho,$  Corollary~\ref{COR:derivation_probability_from_dataset} implies that $(\rhd,Pr_{\rho,\rhd})$ is a justification by compromise of $\rho.$
Since $\vert X^*\vert=1$, let $i\in\{0,\cdots,\vert X\vert-1\}$ be the index such that $\rho(x^{\rhd}_{i+1},X)=1.$
Corollary~\ref{COR:identification_distortions_probability} yields $Pr(\rhd_{i})=1$.
By Corollary~\ref{LEM:uniqueness_up_to_probability_distribution} it is enough to show that for any $j\in\{0,\cdots,\vert X\vert-1\}$ there is $\rhd^{\prime}\in\mathsf{LO}(X)$ such that $\rhd_{i}\equiv \rhd^{\prime}_{j}.$
Consider some $j\in\{0,\cdots,\vert X\vert-1\}.$ 
Let $\rhd^{\prime}$ be defined by 
\begin{align*}
	x^{\rhd^{\prime}}_{k}=x^{\rhd_i}_{\vert X\vert-k+1}  \,\text{for any}\, k\in \{1,\cdots,j\}, 	\,\text{and}\\
 x^{\rhd^{\prime}}_{k}=x^{\rhd_i}_{k+j} \,\text{for any}\, k\in\{j+1,\cdots,\vert X\vert\}.
\end{align*}
{For any $j\in\{0,\cdots,\vert X\vert-1\}$, each $\rhd^{\prime}\in\mathsf{LO}(X)$ is built so that $\rhd^{\prime}_j\equiv \rhd_i.$
 Note also that, for $j=i$ I have that $\rhd^{\prime}=\rhd$.}
Thus, I apply Definition~\ref{DEF:compromise} to conclude that $\rhd^{\prime}_j\equiv
\rhd_{i}.$  
\qed
\smallskip

\noindent \textbf{\large Proof of Theorem~\ref
{THM:computation_stochastic_degree_self_punishment}}.
Let $\rho\colon X\times \X\to [0,1]$ be a compromise-based RUM defined on a ground set of cardinality $\vert X \vert\geq 3$.
Assume that $\vert X^{*}\vert=1.$
Since $\rho$ is compromise-based, by Theorem~\ref{THM:compromise-based_stochastic_choices_characterization} I know that there is $\rhd\in\mathsf{LO}(X)$ that composes $\rho$.
Then by Lemma~\ref{LEMMA:unique_two_justifications_one_item} I obtain that for any $j\in\{0,\cdots,\vert X\vert -1\}$ there is $\rhd^{\prime}\in\mathsf{LO}(X)$ such that $(\rhd^{\prime},Pr_{\rho,\rhd^{\prime}})$ justifies $\rho$ by compromise and $Pr_{\rho,\rhd^{\prime}}(\rhd^{\prime}_j)=1$.
Definition~\ref{DEF:stochastic_degree_self_punishment} implies that $cp(\rho)=0.$

Assume now that $\vert X^*\vert\geq 2$.
First note, that if $cp(\rho)=i,$ 
Definition~\ref{DEF:stochastic_degree_self_punishment} implies that there is a justification by compromise $(\rhd,Pr)$ of $\rho$ such that $Pr(\rhd_i)>0$, and $Pr(\rhd_j)=0$, for any $i<j\leq\vert X\vert-1$	.
Apply Corollary~\ref{COR:derivation_probability_from_dataset} to conclude that $\rho$ has a $(i+1)$-th ordered composition.   

We are left to show that, if $\rho$ has a $i+1$-th ordered composition, then $cp(\rho)=i.$
By Definition~\ref{DEF:ordered_composition_degree_j} there is $\rhd\in\mathsf{LO}(X)$ that composes $\rho,$ $\rho\left(x^{\rhd}_{i+1},X\right)> 0$, and $\rho\left(x^{\rhd}_{l},X\right)=0$ for any $ i+1<l\leq \vert X\vert$. 
Two cases are possible:
\begin{itemize}
	\item[(i)] $\vert X^*\vert=2$, or 
	\item[(ii)]$\vert X^*\vert>2.$
\end{itemize} 
If (i) holds, then without loss of generality there is $h\in\{1,\cdots,\vert X\vert\}$ such that $h\leq i$, $Pr(X,x^{\rhd}_h)>0$, $Pr(X,x^{\rhd}_{i+1})>0$, $Pr(X,x^{\rhd}_h)+Pr(X,x^{\rhd}_{i+1})=1$. 
We must consider two subcases:
\begin{itemize}
\item[(i)(a)] $\min(X,\rhd)\in\vert X^*\vert,$ equivalently $i+1=\vert X\vert$, or
\item[(i)(b)] $\min(X,\rhd)\not\in\vert X^*\vert,$ equivalently $i+1<\vert X\vert.$
\end{itemize}
 
If case (i)(a) holds, by Lemma~\ref{LEMMA:unique_two_justifications_two_items} I know that $(\rhd, Pr_{\rho,\rhd})$ and $(\rhd^{*h},Pr_{\rho,\rhd^{*h}})$ are the only two distinct justifications by compromise of $\rho,$ $Pr_{\rho,\rhd}(\rhd_{\vert X\vert-1})> 0,$ and $Pr_{\rho,\rhd^{*h}}\left(\rhd^{*h}_{\vert X\vert-1}\right)> 0.$
Definition~\ref{DEF:stochastic_degree_self_punishment} implies that $cp(\rho)=\vert X\vert-1=i.$ 

If case (i)(b) holds, then by Theorem~\ref{THM:stochastic_self_punishment_identification} $(\rhd,Pr_{\rho,\rhd})$ is the unique justification by compromise of $\rho.$
Moreover, by Corollary~\ref{COR:identification_distortions_probability} I obtain that $Pr_{\rho,\rhd}(\rhd_{i})>0,$ and  $Pr_{\rho,\rhd}(\rhd_{l})=0,$  for any $i<l\leq\vert X\vert-1.$
Definition~\ref{DEF:stochastic_degree_self_punishment} implies that $cp(\rho)=i.$

If case (ii) holds, then Theorem~\ref{THM:stochastic_self_punishment_identification} implies that $(\rhd,Pr_{\rho,\rhd})$ is the unique justification by compromise of $\rho.$
Moreover, by Corollary~\ref{COR:identification_distortions_probability} I obtain that $Pr_{\rho,\rhd}(\rhd_{i})>0,$ and  $Pr_{\rho,\rhd}(\rhd_{l})=0,$  for any $i<l\leq\vert X\vert-1.$
Definition~\ref{DEF:stochastic_degree_self_punishment} implies that $cp(\rho)=i.$
\qed

\medskip

\noindent \textbf{\large Proof of Lemma~\ref{LEMMA:degree_self_punishment_exact}}.
Let $\rho\colon X\times \X\to [0,1]$ be a compromise-based RUM defined on a ground set of cardinality $\vert X \vert\geq 3$, and such that $\vert X^{*}\vert\geq 2.$
Since $\rho$ is a compromise-based RUM, by Theorem~\ref{THM:compromise-based_stochastic_choices_characterization} there is $\rhd\in\mathsf{LO}(X)$ that composes $\rho$.
Two cases are possible:
\begin{enumerate}[\rm(i)] 
\item $\vert X^*\vert >3,$ or $\vert X^{*}\vert=2$ and $\min(X,\rhd){\not\in} X^*;$ 
\item $\vert X^*\vert=2$ and $\min(X,\rhd)\in X^*$.
\end{enumerate} 

If (i) holds, by Theorem~\ref{THM:stochastic_self_punishment_identification} $\left(\rhd,Pr_{\rho,\rhd}\right)$ is the unique justification by compromise.
Definition~\ref{DEF:Stochastic_lack_of_confidence} yields the claim.
If (ii) holds, let $j\in\{1,\cdots,\vert X\vert-1\}$ be the other index such that $\rho(x^{\rhd}_{j},X)>0.$
By Lemma \ref{LEMMA:unique_two_justifications_two_items} $\left(\rhd,Pr_{\rho,\rhd}\right)$ and $(\rhd^{*j}, Pr_{\rho,\rhd^{*j}})$ are the only two justifications by compromise of $\rho,$  and $Pr_{\rho,\rhd}(\rhd_{j-1})=Pr_{\rho,\rhd^{*j}}\left(\rhd^{*j}_{\vert X\vert-1}\right)> 0,$ $Pr_{\rho,\rhd}(\rhd_{\vert X\vert-1})=Pr_{\rho,\rhd^{*j}}\left(\rhd^{*j}_{j-1}\right)> 0.$
Definition~\ref{DEF:Stochastic_lack_of_confidence} yields the claim again.\qed

\end{document}